\documentclass[aps,prd,superscriptaddress,nofootinbib,tighten,preprint]{revtex4}
\pdfoutput=1
\usepackage[utf8]{inputenc}
\usepackage[english]{babel}
\usepackage{amsmath}
\usepackage{subfigure}
\usepackage{hyperref}
\usepackage{amsfonts}
\usepackage{amssymb}
\usepackage{graphicx}
\usepackage{slashed}
\newcommand{\be}{\begin{equation}}
\newcommand{\ee}{\end{equation}}
\newcommand{\bea}{\begin{eqnarray}}
\newcommand{\eea}{\end{eqnarray}}

\newcommand{\umt}{{\rm U(1)}_{L_{\mu}-L_{\tau}}}
\newcommand{\vmt}{v_{\mu \tau}}
\newcommand{\gmt}{g_{\mu \tau}}
\newcommand{\zmt}{Z_{\mu \tau}}
\newcommand{\mzmt}{M_{\zmt}}
\newcommand{\dm}{\phi_{DM}}
\newcommand{\dmd}{\phi^\dagger_{DM}}

\newcommand{\mdm}{M_{DM}}
\newcommand{\ldh}{\lambda_{Dh}}
\newcommand{\ldH}{\lambda_{DH}}
\newcommand{\nmt}{n_{\mu \tau}}

\newcommand{\smgauge}{{\rm SU}(2)_{\rm L}\times {\rm U}(1)_{\rm Y}}
\newcommand{\nn}{\nonumber}

\catcode`@=12
\def\la{\mathrel{\mathchoice {\vcenter{\offinterlineskip\halign{\hfil
$\displaystyle##$\hfil\cr<\cr\sim\cr}}}
{\vcenter{\offinterlineskip\halign{\hfil$\textstyle##$\hfil\cr<\cr\sim\cr}}}
{\vcenter{\offinterlineskip\halign{\hfil$\scriptstyle##$\hfil\cr<\cr\sim\cr}}}
{\vcenter{\offinterlineskip\halign{\hfil$\scriptscriptstyle##$\hfil\cr<\cr\sim
\cr}}}}}

\begin{document}
\title{FIMP and Muon ($g-2$) in a U$(1)_{L_{\mu}-L_{\tau}}$ Model}

\author{Anirban Biswas}
\email{anirbanbiswas@hri.res.in}
\affiliation{Harish-Chandra Research Institute, Chhatnag Road,
Jhunsi, Allahabad 211 019, India}
\affiliation{Homi Bhabha National Institute,
Training School Complex, Anushaktinagar, Mumbai - 400094, India}
\author{Sandhya Choubey}
\email{sandhya@hri.res.in}
\affiliation{Harish-Chandra Research Institute, Chhatnag Road,
Jhunsi, Allahabad 211 019, India}
\affiliation{Homi Bhabha National Institute,
Training School Complex, Anushaktinagar, Mumbai - 400094, India}
\affiliation{Department of Theoretical Physics, School of
Engineering Sciences, KTH Royal Institute of Technology, AlbaNova
University Center, 106 91 Stockholm, Sweden}
\author{Sarif Khan}
\email{sarifkhan@hri.res.in}
\affiliation{Harish-Chandra Research Institute, Chhatnag Road,
Jhunsi, Allahabad 211 019, India}
\affiliation{Homi Bhabha National Institute,
Training School Complex, Anushaktinagar, Mumbai - 400094, India}

\begin{abstract} 
The tightening of the constraints on the standard thermal WIMP scenario 
has forced physicists to propose alternative dark matter (DM) models.
One of the most popular alternate explanations of the origin of DM is
the non-thermal production of DM via freeze-in. In this scenario
the DM never attains thermal equilibrium with the thermal soup 
because of its feeble coupling strength ($\sim 10^{-12}$)
with the other particles in the thermal bath and is generally
called the Feebly Interacting Massive Particle (FIMP).
In this work, we present a gauged U(1)$_{L_{\mu}-L_{\tau}}$
extension of the Standard Model (SM) which has a scalar
FIMP DM candidate and can consistently explain the DM relic density 
bound. In addition, the spontaneous breaking of the
U(1)$_{L_{\mu}-L_{\tau}}$ gauge symmetry gives an extra
massive neutral gauge boson $Z_{\mu\tau}$ which can
explain the muon ($g-2$) data through its additional one-loop 
contribution to the process. Lastly, presence of three right-handed 
neutrinos enable the model to successfully explain the small neutrino 
masses via the Type-I seesaw mechanism. The presence of the 
spontaneously broken U(1)$_{L_{\mu}-L_{\tau}}$ gives a particular 
structure to the light neutrino mass matrix which can explain the 
peculiar mixing pattern of the light neutrinos. 
\end{abstract}
\maketitle
\newpage
\section{Introduction}
The presence of Dark Matter (DM) is a well established truth and 
demands the extension of the Standard Model (SM). 
Starting from the time of Zwicky (who examined the coma-cluster in 1933
and gave the prediction of excess matter) till now, many
collaborations have reported  
the existence of DM. The most prominent examples are, flatness
of galaxy rotation curve \cite{Sofue:2000jx},
gravitational lensing \cite{Bartelmann:1999yn} and the
Bullet cluster observed by NASA's Chandra satellite
\cite{Clowe:2003tk}. In the last one decade,
satellite borne experiments like WMAP
(Wilkinson Massive Anisotropy Probe) \cite{Hinshaw:2012aka} and 
Planck \cite{Ade:2015xua} not only confirmed the existence
of DM but also measured the amount of DM present in the Universe
with an unprecedented accuracy. However, the particle content
of dark matter as well as its interaction strength
with the visible world are two questions which
remain unresolved. Depending
on their production mechanism in the early Universe, the dark matter
particles can be classified into two groups. The first group is  
the thermal dark matter. These
class of particles were produced thermally
and maintained their thermal as well as
chemical equilibrium with the thermal bath.
They eventually decoupled from the plasma
when their interaction rates became less than
the expansion rate of the Universe. Being
{\it relic} particles, their comoving
number density froze to a particular value. This
is known as the usual freeze-out mechanism
\cite{Gondolo:1990dk, Srednicki:1988ce}.
The dark matter candidates which were produced 
through the freeze-out scenario are usually called the 
Weakly Interacting Massive Particle
or WIMP \cite{Gondolo:1990dk, Srednicki:1988ce}. These types of particles
have weak scale interaction cross section
and their signatures are expected to be
seen at various ongoing direct 
detection experiments like LUX \cite{Akerib:2016vxi, Akerib:2015rjg} and 
XENON-1T \cite{Aprile:2015uzo}. However, 
the direct detection experiments are yet to  
find any {\it real signal} of
dark matter particles. This forces us to look for
other types of dark matter particles
which can be an alternative to the WIMP scenario.
One such class of dark matter candidates is the 
Feebly Interacting Massive Particle
or FIMP \cite{Hall:2009bx, Yaguna:2011qn, Molinaro:2014lfa,
Biswas:2015sva, Merle:2015oja, Shakya:2015xnx, Biswas:2016bfo,
Konig:2016dzg}. The interaction rates of these particles
are so feeble that they never attained thermal equilibrium
with the thermal bath. As a result, their initial
number densities were negligible compared
to others which were in thermal equilibrium.
As the Universe cooled down, they were produced
predominantly from the decays of heavy particles. 
If the decaying mother particles are in
thermal equilibrium at the early stage of the Universe, the production
of dark matter particles from a particular decay mode is expected to
be maximum around a temperature equal to the mass of the corresponding
decaying particle. In principle, FIMP can also be produced from
the annihilation of SM particles and other heavy
particles (beyond SM) as well. The production mechanism of FIMP
is known as freeze-in \cite{Hall:2009bx} which is opposite to
the freeze-out scenario. Unlike the freeze-out mechanism,
in case of freeze-in the comoving number density and hence
the relic density of a FIMP is directly proportional to its
couplings with other particles. Since the interaction
strength of a FIMP is extremely weak, thus one can evade
all the constraints arising from the direct detection experiments.

Besides the dark sector,
the other puzzles which require the extension of the SM are the 
observation of  
neutrino mass \cite{Ahmad:2002jz}, discrepancy in
the anomalous magnetic moment of muon ($g-2$) \cite{Bennett:2006fi}
from its SM prediction and also
the baryon asymmetry of the Universe \cite{Fukugita:1986hr}.
The existence of extremely small neutrino masses
were first confirmed by the Super-Kamiokande 
collaboration \cite{Fukuda:1998mi} from the
observation of neutrino flavor oscillations
in atmospheric neutrino data.
Thereafter, many outstanding neutrino
experiments like SNO \cite{Ahmad:2002jz}, 
KamLand \cite{Eguchi:2002dm},
Daya Bay \cite{An:2015nua}, RENO \cite{RENO:2015ksa},
Double Chooz \cite{Abe:2014bwa},  
T2K \cite{Abe:2015awa, Salzgeber:2015gua} and NO$\nu$A
\cite{Adamson:2016tbq, Adamson:2016xxw} have precisely
measured the two mass squared differences and
three mixing angles between different neutrino flavours.

In the present work, we try to explain the existence of dark matter,
tiny nature of neutrino masses, their intergenerational
mixing angles and muon ($g-2$) anomaly simultaneously
within the framework of our proposed model. 
Since we already know that the SM is unable
to address these issues, we have to think of a
beyond Standard Model (BSM) scenario.
There are many well motivated extensions of the 
SM like SUSY, 
two Higgs doublet model,
extension of the SM 
gauge group by extra U(1)
and many more. In this work we have extended the
SM gauge group by a local $\umt$ symmetry
\cite{Choubey:2004hn,Biswas:2016yan,Altmannshofer:2016jzy}.
Therefore, the complete gauge group in this model is 
$\smgauge \times \umt$. 
Since we are increasing
the SM gauge group by a local $\umt$ symmetry,
we must check the cancellation
of axial vector anomaly \cite{Adler:1969gk, Bardeen:1969md} and 
mixed gravitational-gauge anomaly \cite{Delbourgo:1972xb, Eguchi:1976db}.
The advantage of the $\umt$ extension
is that these anomalies cancel automatically between the 
second and third generations \cite{He:1990pn,
He:1991qd, Ma:2001md} of SM fermions.

We also extend the particle content of the SM and include three right-handed 
neutrinos and two SM gauge singlet scalars. The new particles are given 
appropriate $\umt$ charge. 
One of the scalars picks up a Vacuum 
Expectation Value (VEV) breaking $\umt$ symmetry spontaneously, while the 
other does not take any VEV. The $\umt$ charge $\nmt$
of the other scalar $\dm$ is 
chosen in such a way that it remains stable even after $\umt$ breaking 
and becomes a dark matter candidate. The new gauge boson $\zmt$
after spontaneous breaking of $\umt$ becomes massive and gives 
additional contribution to the muon $(g-2)$, helping reconcile the observed 
data with the theoretical prediction in this model. 
The three right-handed neutrinos carry Majorana
masses and give rise to light neutrino mass matrix
via the Type-I seesaw mechanism.  
In our earlier work in Ref. \cite{Biswas:2016yan}, we
showed that our model can explain the nonzero
neutrino masses and their intergenerational
mixing angles. Also in that work, we considered
the scalar field $\dm$ as a WIMP type dark matter
candidate and checked its viability in
various direct detection experiments. We
found that all the existing constraints are
satisfied around the resonance regions,
where the mediator mass is nearly equal to
twice of dark matter mass. 

In this work, we show that $\dm$ could serve as a FIMP type dark
matter candidate in this $\umt$ extension of the SM. 
All particles of our model including the additional scalar $h_2$, 
$\zmt$ and the three right-handed neutrinos 
are in thermal equilibrium in the early Universe except $\dm$. 
In order to make this possible, 
$\dm$ must be feebly interacting with other particles. 
We choose its couplings with the visible
sector to be extremely small (see Section \ref{model} and \ref{b-euation} for
detailed discussions) such that it remains out of equilibrium
throughout its evolution in the early Universe \footnote{$\Gamma<H$
where $\Gamma$ is the dark matter production rate and $H$ is the Hubble parameter
\cite{Arcadi:2013aba}}. We compute the relic density of
$\dm$ by solving the Boltzmann equation where we consider
all the possible production modes of $\dm$ from the decays
as well as the annihilations of SM and BSM particles.
We find that in our model, $\dm$ is
produced not only from the decays of $h_1$ and $h_2$, but also from the pair
annihilation of $N_i$ ($i=2$, 3) mediated by the extra neutral gauge
boson $\zmt$. Therefore, the dark matter phenomenology is intricately 
intertwined with the phenomenology of neutrino masses and muon $(g-2)$.

Rest of the paper is organised in the following way.
In Section \ref{model} we describe the model in detail.
In Section \ref{muon g-2 neutrino mass} we discuss  
muon ($g-2$) and neutrino masses and mixings. 
The Boltzmann equation for FIMP considering all possible
production channels is given in Section \ref{b-euation}. 
Our main results are presented in Section \ref{res}.
Finally, we summarise the present work in Section \ref{con}. All the relevant
decay widths and annihilation cross sections are given in 
Appendices A and B.     

\section{Model}
\label{model}

We have considered the $\umt$ extension of the SM, where
$L_{\mu}$ and $L_{\tau}$ are the muon and tau lepton numbers, respectively.
Hence, the complete gauge group of the minimally extended SM is $\smgauge \times \umt$. 
The particle content of our model is given in Tables \ref{tab1} and \ref{tab2}. 
\begin{center}
\begin{table}[h!]
\begin{tabular}{||c|c|c|c||}
\hline
\hline
\begin{tabular}{c}
    Gauge\\
    Group\\ 
    \hline
    
    ${\rm SU(2)}_{\rm L}$\\ 
    \hline
    ${\rm U(1)}_{\rm Y}$\\ 
\end{tabular}
&

\begin{tabular}{c|c|c}
    \multicolumn{3}{c}{Baryon Fields}\\ 
    \hline
    $Q_{L}^{i}=(u_{L}^{i},d_{L}^{i})^{T}$&$u_{R}^{i}$&$d_{R}^{i}$\\ 
    \hline
    $2$&$1$&$1$\\ 
    \hline
    $1/6$&$2/3$&$-1/3$\\ 
\end{tabular}
&
\begin{tabular}{c|c|c}
    \multicolumn{3}{c}{Lepton Fields}\\
    \hline
    $L_{L}^{i}=(\nu_{L}^{i},e_{L}^{i})^{T}$ & $e_{R}^{i}$ & $N_{R}^{i}$\\
    \hline
    $2$&$1$&$1$\\
    \hline
    $-1/2$&$-1$&$0$\\
\end{tabular}
&
\begin{tabular}{c|c|c}
    \multicolumn{3}{c}{Scalar Fields}\\
    \hline
    $\phi_{h}$&$\phi_{H}$&$\phi_{DM}$\\
    \hline
    $2$&$1$&$1$\\
    \hline
    $1/2$&$0$&$0$\\
\end{tabular}\\
\hline
\hline
\end{tabular}
\caption{Charges of all particles under SM gauge group.}
\label{tab1}
\end{table}
\end{center}
\begin{center}
\begin{table}[h!]
\begin{tabular}{||c|c|c|c||}
\hline
\hline
\begin{tabular}{c}
    Gauge\\
    Group\\ 
    \hline
    $\umt$\\ 
    
\end{tabular}
&
\begin{tabular}{c}
    \multicolumn{1}{c}{Baryonic Fields}\\ 
    \hline
    $(Q^{i}_{L}, u^{i}_{R}, d^{i}_{R})$\\ 
    \hline
    $0$ \\ 
    
\end{tabular}
&
\begin{tabular}{c|c|c}
    \multicolumn{3}{c}{Lepton Fields}\\ 
    \hline
    $(L_{L}^{e}, e_{R}, N_{R}^{e})$ & $(L_{L}^{\mu}, \mu_{R},
    N_{R}^{\mu})$ & $(L_{L}^{\tau}, \tau_{R}, N_{R}^{\tau})$\\ 
    \hline
    $0$ & $1$ & $-1$\\ 
    
\end{tabular}
&
\begin{tabular}{c|c|c}
    \multicolumn{3}{c}{Scalar Fields}\\
    \hline
    $\phi_{h}$ & $\phi_{H}$ & $\phi_{DM}$ \\
    \hline
    $0$ & $1$ & $n_{\mu \tau}$\\
\end{tabular}\\
\hline
\hline
\end{tabular}
\caption{Charges of all particles under $\umt$ gauge group.}
\label{tab2}
\end{table}
\end{center}  
The Lagrangian for this model is as follows,
\begin{eqnarray}
\mathcal{L}&=&\mathcal{L}_{SM} + \mathcal{L}_{N} + \mathcal{L}_{DM}
+ (D_{\mu}\phi_{H})^{\dagger} (D^{\mu}\phi_{H})
-V(\phi_{h},\phi_{H})
-\frac{1}{4} F_{\mu \tau}^{\alpha \beta} {F_{\mu \tau}}_{\alpha \beta}
\,, 
\label{lag}
\end{eqnarray}
where $\mathcal{L}_{SM}$ represents the Lagrangian of SM and $\mathcal{L}_{N}$
is the Lagrangian for the RH neutrinos which contains its kinetic energy terms,
mass terms and Yukawa terms with the SM leptons doublets. The expression of
$\mathcal{L}_{N}$ is,
\begin{eqnarray}
\mathcal{L}_{N}&=&
\sum_{\alpha=e,\,\mu,\,\tau}\frac{i}{2}
\bar{N_\alpha}\gamma^{\mu}D_{\mu}N_{\alpha} 
-\dfrac{1}{2}\,M_{ee}\,\bar{N_e^{c}}N_{e}
-\dfrac{1}{2}\,M_{\mu \tau}\,(\bar{N_{\mu}^{c}}N_{\tau}
+\bar{N_{\tau}^{c}}N_{\mu})  \nn \\ &&
-\dfrac{1}{2}\,h_{e \mu}(\bar{N_{e}^{c}}N_{\mu} 
+\bar{N_{\mu}^{c}}N_{e})\phi_H^\dagger
- \dfrac{1}{2}\,h_{e \tau}(\bar{N_{e}^{c}}N_{\tau} 
+ \bar{N_{\tau}^{c}}N_{e})\phi_H
\nn \\ &&
-\sum_{\alpha=e,\,\mu,\,\tau} y_{\alpha} \bar{L_{\alpha}}
\tilde {\phi_{h}} N_{\alpha} +h.c.\,
\label{lagN}
\end{eqnarray}  
where $\tilde {\phi_{h}}=i\,\sigma_2\phi^*_h$ and 
the quantities $M_{\mu \tau}$, $M_{ee}$ have dimension of mass while the
Yukawa couplings $h_{e\mu}$, $h_{e \tau}$
and $y_{\alpha}$s are dimensionless constants.
In Eq.\,(\ref{lag}), $\mathcal{L}_{DM}$
part is the dark matter Lagrangian which contains the kinetic term of the dark matter candidate
$\phi_{DM}$ and the interaction terms of $\phi_{DM}$ with the SM-like
Higgs ($\phi_{h}$) and the extra Higgs ($\phi_{H}$). The expression of the
$\mathcal{L}_{DM}$ is as follows,
\begin{eqnarray}
\mathcal{L}_{DM} &=& (D^{\mu}\phi_{DM})^\dagger (D_{\mu}\phi_{DM})
- \mu_{DM}^{2} \phi_{DM}^{\dagger} \phi_{DM} 
-\lambda_{DM} (\phi_{DM}^{\dagger} \phi_{DM})^{2}
\nn \\ && 
-\lambda_{Dh}(\phi_{DM}^{\dagger} \phi_{DM})
(\phi_{h}^{\dagger} \phi_{h}) 
-\lambda_{DH}(\phi_{DM}^{\dagger} \phi_{DM})
(\phi_{H}^{\dagger} \phi_{H})\,.
\label{ldm}
\end{eqnarray}
The fourth term in Eq.\,(\ref{lag}) represents
the kinetic term of the extra Higgs singlet $\phi_{H}$. The potential
$V(\phi_{h},\phi_{H})$ contains the quadratic and quartic interactions of
$\phi_{H}$ and the interaction term between $\phi_{h}$ and $\phi_{H}$. The potential
$V(\phi_{h},\phi_{H})$ has the following form,
\begin{eqnarray}
V(\phi_h, \phi_H) = \mu_{H}^{2} \phi_{H}^{\dagger} \phi_{H} 
+ \lambda_{H} (\phi_{H}^{\dagger} \phi_{H})^{2}
+ \lambda_{hH}(\phi_{h}^{\dagger} \phi_{h}) (\phi_{H}^{\dagger} \phi_{H}) \,.
\label{int}
\end{eqnarray} 
The last term in the Eq.\,(\ref{lag}) is the kinetic energy term of the extra
gauge boson Z$_{\mu\tau}$ in terms of the field strength tensor
$F^{\alpha\beta}_{\mu\tau}$ which has the following form,
$F_{\mu \tau}^{\alpha \beta} = \partial^\alpha \zmt^\beta
-\partial^\beta \zmt^\alpha.$ We can write down the generic form of the covariant
derivative which are appearing in the Eqs.\,(\ref{lag})-(\ref{ldm}) is,
\begin{eqnarray}
D_{\nu} X = (\partial_{\nu} + i\,\gmt\,Q_{\mu \tau} (X)\,{\zmt}_{\nu})\,X\,,
\end{eqnarray}
where $X$ is any SM singlet field which has $\umt$ charge $Q_{\mu\tau}(X)$
(see Table \ref{tab2}) and $g_{\mu\tau}$ is the gauge coupling of
the $\umt$ gauge group.
The $\umt$ symmetry breaks
spontaneously once $\phi_{H}$ picks up a VEV 
and consequently the extra gauge field becomes
massive with mass term  $M_{\zmt}=\gmt\,\vmt$. In the unitary
gauge, $\phi_{h}$ and $\phi_{H}$ take the following form after
spontaneous breaking of the  $\smgauge\times\umt$ gauge symmetry,
\begin{eqnarray}
\phi_{h}=
\begin{pmatrix}
0 \\
\dfrac{v+H}{\sqrt{2}}
\end{pmatrix}\,,
\,\,\,\,\,\,\,\,\,
\phi_{H}=
\begin{pmatrix}
\dfrac{\vmt + H_{\mu\tau}}{\sqrt{2}}
\end{pmatrix}\,,
\label{phih}
\end{eqnarray}
where $v$ and $\vmt$ are the VEVs of $\phi_{h}$ and $\phi_{H}$
respectively. The presence of the mutual interaction
between $\phi_{h}$ and $\phi_{H}$ in Eq.~(\ref{int}) induces
the mixing between $H$ and $H_{\mu\tau}$. Hence, the scalar mixing
matrix has the following form,
\begin{eqnarray}
\mathcal{M}^2_{scalar} = \left(\begin{array}{cc}
2\lambda_h\,v^2 ~~&~~ \lambda_{hH}\,\vmt\,v \\
~~&~~\\
\lambda_{hH}\,\vmt\,v ~~&~~ 2 \lambda_H\,\vmt^2
\end{array}\right) \,\,.
\label{mass-matrix}
\end{eqnarray}
We see from the expression of $\mathcal{M}^2_{scalar}$ that if we take
$\lambda_{hH}=0$, then $H$ and $H_{\mu\tau}$ can represent the
physical states (zero mixing between $H$ and $H_{\mu\tau}$).
But in our case, $\lambda_{hH} \neq 0$ and therefore,
we need to introduce two physical states in the following way,
\begin{eqnarray}
h_{1}&=& H \cos \alpha + H_{\mu\tau} \sin \alpha \,, \nn \\
h_{2}&=& - H \sin \alpha + H_{\mu\tau} \cos \alpha\,,
\end{eqnarray}
where $\alpha$ is the mixing angle. The above mixing matrix
becomes diagonal with respect to the physical states $h_1$,
$h_2$ and the eigenvalues of $\mathcal{M}^2_{scalar}$ represent
the mass terms $M_{h_1}$
and $M_{h_2}$ of the scalar fields $h_1$ and $h_2$ respectively.
In this work, we choose $h_1$ as the SM-like Higgs boson. 
The quartic couplings
$\lambda_{h}$, $\lambda_{H}$ and $\lambda_{hH}$ 
can be expressed in terms of $M_{h_1}$, $M_{h_2}$, VEVs
($v$, $\vmt$) and mixing angle ($\alpha$) in the following way,
\begin{eqnarray}
\lambda_{H} &=& \dfrac{M_{h_{2}}^{2} + M_{h_{1}}^{2} -
(M_{h_{1}}^{2} - M_{h_{2}}^{2})\cos 2 \alpha}{4\,v_{\mu\tau}^{2}}\,,\nn \\
\lambda_{h}&=& \dfrac{M_{h_{2}}^{2} + M_{h_{1}}^{2} -
(M_{h_{2}}^{2} - M_{h_{1}}^{2})\cos 2 \alpha}{4\,v^{2}}\,,\nn\\
\lambda_{hH} &=& -\dfrac{(M_{h_{2}}^{2}-M_{h_{1}}^{2})
\cos \alpha \sin \alpha}{v\,v_{\mu\tau}}\,.
\label{lamhH}
\end{eqnarray}
After symmetry breaking, mass of the dark matter is,
\begin{eqnarray}
\mdm^2 = \mu^2_{DM} + \dfrac{\lambda_{Dh}\,v^2}{2} +
\dfrac{\lambda_{DH}\,\vmt^2}{2}\,.
\end{eqnarray}
In the present scenario,
two of the three scalars take VEVs while the third one does not,  i.e.,
$\langle\phi_{h}\rangle = \dfrac{v}{\sqrt{2}}$,
$\langle\phi_{H}\rangle = \dfrac{\vmt}{\sqrt{2}}$ and
$\langle\dm\rangle = 0$. This requires
\begin{eqnarray}
\mu^2_h<0,\,\,\,\mu^2_H<0\,\,\text{and}\,\,\,\mu^2_{DM}>0.
\end{eqnarray} 
In addition, the stability of the vacuum requires \cite{Biswas:2016ewm} 
the following constraints on the quartic couplings of the
scalar fields,
\begin{eqnarray}
&&\lambda_h \geq 0, \lambda_H \geq 0, \lambda_{DM} \geq 0,\nonumber \\
&&\lambda_{hH} \geq - 2\sqrt{\lambda_h\,\lambda_H},\nonumber \\
&&\lambda_{Dh} \geq - 2\sqrt{\lambda_h\,\lambda_{DM}},\nonumber \\
&&\lambda_{DH} \geq - 2\sqrt{\lambda_H\,\lambda_{DM}},\nonumber \\
&&\sqrt{\lambda_{hH}+2\sqrt{\lambda_h\,\lambda_H}}\sqrt{\lambda_{Dh}+
2\sqrt{\lambda_h\,\lambda_{DM}}}
\sqrt{\lambda_{DH}+2\sqrt{\lambda_H\,\lambda_{DM}}} \nonumber \\ 
&&+ 2\,\sqrt{\lambda_h \lambda_H \lambda_{DM}} + \lambda_{hH} \sqrt{\lambda_{DM}}
+ \lambda_{Dh} \sqrt{\lambda_H} + \lambda_{DH} \sqrt{\lambda_h} \geq 0 \,\,\,\,.
\end{eqnarray}
Besides the lower limit, there is an upper limit on the
quartic couplings and the Yukawa couplings due to the perturbative regime,
constraining them to be less than $4\pi$ and $\sqrt{4\pi}$, respectively. 

\section{Muon $(g-2)$ and neutrino mass}
\label{muon g-2 neutrino mass}

In our previous work \cite{Biswas:2016yan}
we studied  in detail the muon ($g-2$) and neutrino mass 
phenomenology in the framework of present model. The difference
here with the previous work is that here we wish to produced the dark matter via 
the freeze-in mechanism instead of freeze-out, as in \cite{Biswas:2016yan}.
Our aim is to simultaneously explain the dark matter relic
abundance as well as the muon ($g-2$).  
The region of the $g_{\mu\tau}$\,-\,$M_{Z_{\mu\tau}}$ 
parameter space that can explain the muon ($g - 2$) has been shown in 
\cite{Altmannshofer:2014pba}. This allowed region is seen to be very small 
and mostly constrained by the neutrino trident process experiments
such as CHARM-II, CCFR \cite{Geiregat:1990gz,Mishra:1991bv}
and 4-leptons decay \cite{Altmannshofer:2014pba}. 
\begin{figure}[h!]
\centering
\includegraphics[angle=0,height=6cm,width=9cm]{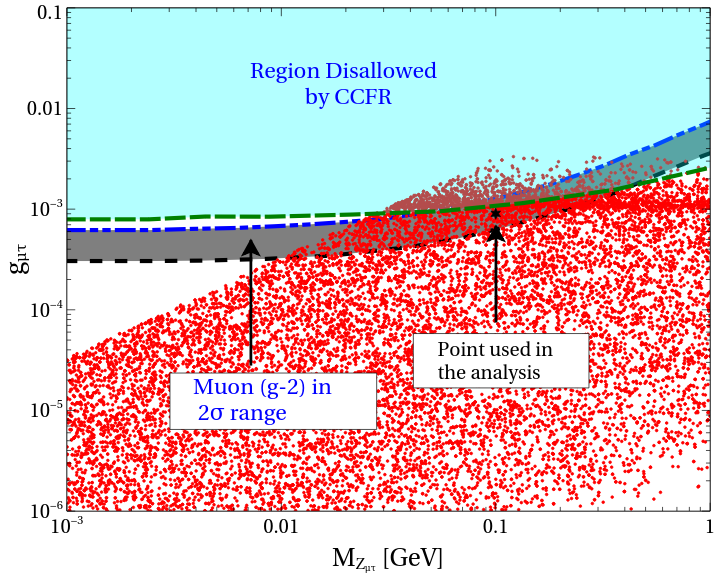}
\caption{Bounds in the $g_{\mu\tau}$\,-\,$M_{Z_{\mu\tau}}$ plane
from different experiments and allowed region to satisfy relic
density (red dots) and muon ($g-2$) excess in 2\,$\sigma$
range (grey shaded region).}
\label{gmt-mzmt}
\end{figure}
In determining the allowed parameter space for the present model,
we have considered the relic density constraint \cite{Ade:2015xua}
and also the bound on invisible decay width
of SM-like Higgs \cite{Khachatryan:2016vau}.
The one loop contribution \cite{Gninenko:2001hx,Baek:2001kca} to muon ($g-2$) 
for the $\umt$ gauge boson Z$_{\mu\tau}$ mediated diagram is given by,
\begin{eqnarray}
\Delta a_{\mu}(Z_{\mu \tau}) = \dfrac{g_{\mu \tau}^{2}}
{8 \pi^{2}} \int_{0}^{1} dx \dfrac{2 x(1-x)^{2}}{(1-x)^{2} + rx}\,,
\label{intg2}
\end{eqnarray}
where, $r = (M_{Z_{\mu \tau}}/m_{\mu})^{2}$, $m_{\mu}$ being the muon mass. 
The discrepancy between the observed and SM predicted 
value \cite{Jegerlehner:2009ry} of muon ($g-2$) is,
\begin{eqnarray}
\Delta a_{\mu} = a_{\mu}^{\rm exp} - a_{\mu}^{\rm th}
= (29.0 \pm 9.0) \times 10^{-10}\,.
\label{delta}
\end{eqnarray}
In Fig.~\ref{gmt-mzmt} we show the allowed and disfavoured regions 
in the $g_{\mu\tau}$-$M_{Z_{\mu\tau}}$ plane. 
The region above the green dashed line is ruled out by the
neutrino trident experiment CCFR \cite{Altmannshofer:2014pba}. 
The grey region inside the blue dashed-dotted line
and black dashed
line can explain the ($g-2$) anomaly 
in $\pm$2$\sigma$ range \cite{Altmannshofer:2014pba}. 
As will be discussed in much detail later, the 
red dots in this figure span the parameter region which can satisfy the 
dark matter relic abundance (cf. Eq.~(\ref{rel-val})). We see that for 
$g_{\mu\tau} \geq 3\times10^{-3}$ no red points exist because the 
contribution from the $Z_{\mu\tau}$ 
mediated diagram to the relic abundance becomes too large. 
See Appendix \ref{App:AppendixA}, for the expression of cross section of 
$\bar{N_{i}}\,N_{i}\rightarrow \phi_{DM}^{\dagger} \phi_{DM}$, $i = 2,3$.
The region of the parameter space compatible with both the
dark matter relic abundance and muon ($g-2$)
lies in the narrow overlapping zone. 
The benchmark point (values of $\gmt$ and $\mzmt$)
used in all further results shown in this work is 
marked by the star in Fig.~\ref{gmt-mzmt} and corresponds to 
$\mzmt=100$ MeV and $\gmt=9 \times 10^{-4}$. Such low mass $\zmt$
gauge boson can be searched by looking $2 \mu + \slashed{E_T}$
final states in LHC or future collider experiments
\cite{Elahi:2015vzh,Harigaya:2013twa}. For these values of 
the parameters, the contribution to muon ($g-2$) from Eq.~(\ref{intg2}) 
comes out to be 
\begin{eqnarray}
\Delta a_{\mu} = 22.6 \times 10^{-10}
\,,
\end{eqnarray}     
which lies within the $\pm$2$\sigma$ range of the observed value
\cite{Altmannshofer:2014pba}.

The neutrino mass generation via the Type-I seesaw has been discussed in 
detail in \cite{Biswas:2016yan}. From the neutrino part of the Lagrangian
given in the Eq.~(\ref{lagN}), we can write down the Majorana mass
matrix  for the RH neutrinos after spontaneous symmetry breaking as,
\begin{eqnarray}
M_{R} = \left(\begin{array}{ccc}
M_{ee} ~~&~~ \dfrac{ \vmt}{\sqrt{2}} h_{e \mu}
~~&~~\dfrac{\vmt}{\sqrt{2}} h_{e \tau} \\
~~&~~\\
\dfrac{\vmt}{\sqrt{2}} h_{e \mu} ~~&~~ 0
~~&~~ M_{\mu \tau} \,e^{i\xi}\\
~~&~~\\
\dfrac{\vmt}{\sqrt{2}} h_{e \tau} ~~&
~~ M_{\mu \tau}\,e^{i\xi} ~~&~~ 0 \\
\end{array}\right) \,.
\label{mncomplex}
\end{eqnarray} 
In general $M_{R}$ \footnote{In order to find the mass eigenstates
$N_j$s ($j=1,\,2,\,3$) we have to diagonalise the matrix $M_{R}$.}
can be complex but by proper phase rotation all
the components can be made real except one and
we choose to take $M_{\mu\tau}$ as the complex 
component \cite{Baek:2015mna}. 
The light neutrino mass matrix is given as,
\begin{eqnarray}
m_{\nu}&\simeq&-M_D\,M^{-1}_R M^T_D\,, 
\label{activemass}
\end{eqnarray}
where $M_{D}$ is the Dirac mass term, and here it comes as diagonal. The
expression for the neutrino mass matrix is,
\begin{eqnarray}
m_{\nu} = \dfrac{1}{2\,p} \left(\begin{array}{ccc}
2\,f_{e}^{2}M_{\mu \tau}^{2}e^{i\xi} &
-\sqrt{2}\,f_{e} f_{\mu}\,h_{e \tau} \vmt &
-\sqrt{2}\,f_{e} f_{\tau}\,h_{e \mu} \vmt\\
-\sqrt{2}\,f_{e} f_{\mu}\,h_{e \tau} \vmt &
\dfrac{f_{\mu}^{2}\,h_{e\tau}^2\,\vmt^2\,e^{-i\xi}}{M_{\mu\tau}} &
\dfrac{f_{\mu}\,f_{\tau}}{M_{\mu\tau}}(M_{ee}\,M_{\mu\tau}-p\,e^{-i\xi})\\
-\sqrt{2}\,f_{e} f_{\tau}\,h_{e \mu} \vmt &
\dfrac{f_{\mu}\,f_{\tau}}{M_{\mu\tau}}(M_{ee}\,M_{\mu\tau}-p\,e^{-i\xi})  &
\dfrac{f_{\tau}^{2}\,h_{e\mu}^2\,\vmt^2\,e^{-i\xi}}{M_{\mu\tau}} \\
\end{array}\right) \,\,,
\label{mass-matrix}
\end{eqnarray}
where $p = h_{e\mu}\,h_{e\tau}\,\vmt^2-M_{ee}\,M_{\mu\tau}\,e^{i\xi}$
and $f_e$, $f_{\mu}$ and $f_{\tau}$ are the diagonal components of $M_{D}$.
In \cite{Biswas:2016yan}, we have shown that for the chosen
($g_{\mu\tau}$, $M_{Z_{\mu\tau}}$) value, we can satisfy all
the constraints of the mixing angles, mass square differences
\cite{Capozzi:2016rtj} and cosmological bound on the sum of
the neutrino masses \cite{Ade:2015xua}. We also showed the region
in the model parameter space which satisfy all the neutrino parameter
constraints. 

\section{FIMP Dark Matter and Boltzmann Equation}
\label{b-euation}

\begin{figure}[h!]
\centering
\includegraphics[angle=0,height=12cm,width=14cm]{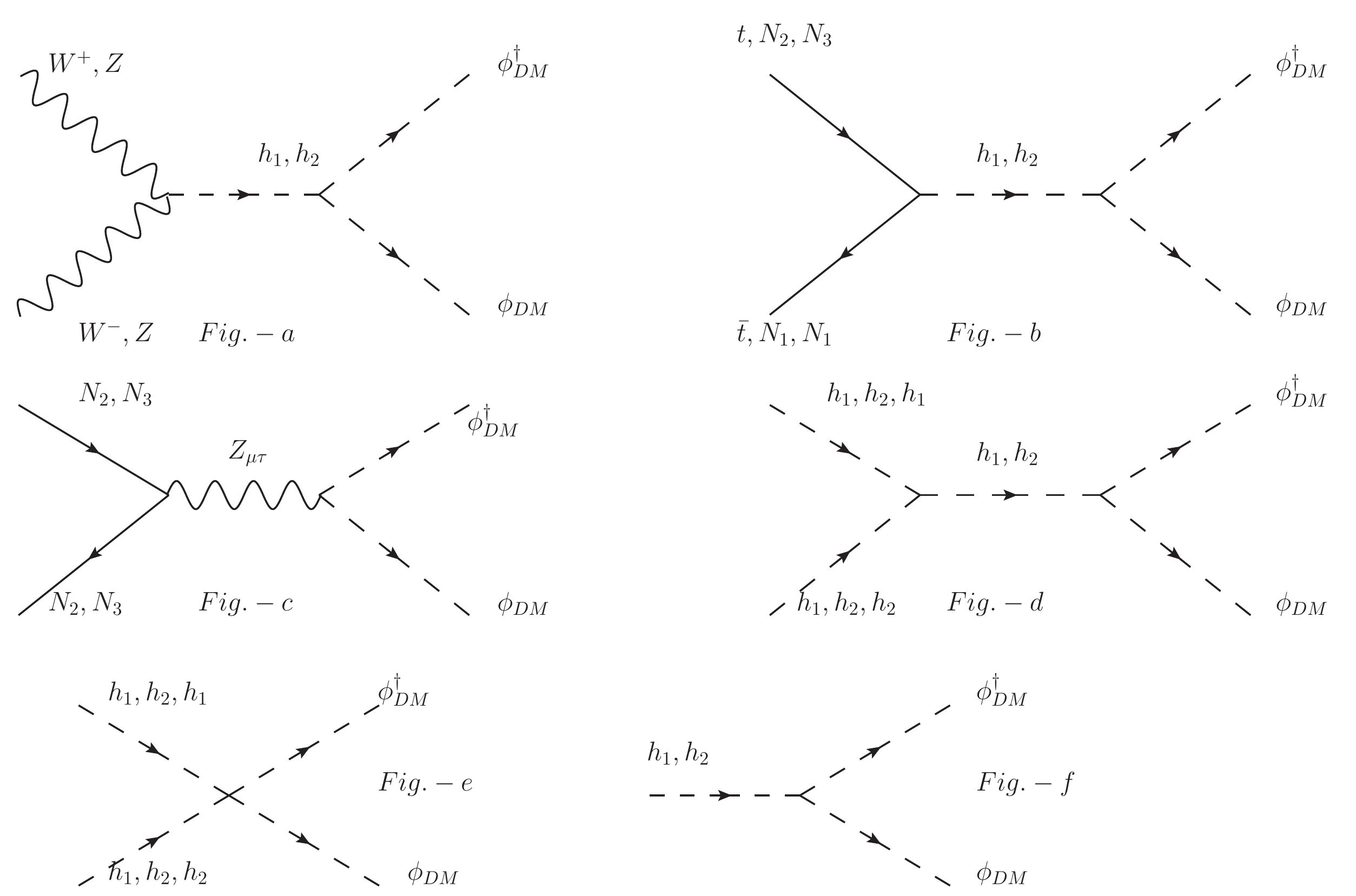}
\caption{Feynman diagrams for the dark matter production processes
from the annihilations and decays of different SM and
BSM particles.}
\label{feyn}
\end{figure}

\begin{table}[h!]
\begin{center}
\vskip 0.5cm
\begin{tabular} {||c||c||}
\hline
\hline
Vertex & Vertex Factor\\
$a\,b\,c$ & $g_{abc}$\\
\hline
$\dm\,\dmd\,h_1$ & $-(\lambda_{Dh} v \cos \alpha + \lambda_{DH} \vmt \sin \alpha)$\\
\hline
$\dm\,\dmd\,h_2$ & $(\lambda_{Dh} v \sin \alpha - \lambda_{DH} \vmt \cos \alpha)$\\
\hline
$\dm\,\dmd\,\zmt^{\rho}$ & $n_{\mu\tau} \gmt (p_2 - p_1)^{\rho}$\\
\hline
$\bar{N_i}\,N_i\,\zmt^{\rho}$ & $\dfrac{\gmt}{2} \gamma^{\rho} \gamma^{5}$ \\
\hline
\hline
\end{tabular}
\end{center}
\caption{Relevant couplings required to compute Feynman
diagrams given in Fig.~\ref{feyn}.}
\label{tab3.1}
\end{table}

As can be seen from the Eqs.~(\ref{lag}--\ref{int}), 
the dark matter particle can interact
with the thermal bath containing both SM as well as BSM particles only
via $h_1$, $h_2$ and $\zmt$. The Feynman diagrams 
relevant for the production of dark matter are shown in Fig.~\ref{feyn}. 
The corresponding couplings are listed
in Table \ref{tab3.1}. From Table \ref{tab3.1}, one can see that the couplings
of $\dm$ with scalar bosons $h_1$ and $h_2$ depend on the parameters
$\ldh$ and $\ldH$, while that with extra neutral gauge boson $\zmt$
involves $\gmt$ and $\nmt$. For the dark matter to be a suitable FIMP 
candidate, the cross section of the diagrams listed in Fig.~\ref{feyn}
should be very small. The complete expressions for the contribution from 
each of the diagrams is given in Appendix A and B. 
The processes involving $h_1$ and $h_2$ 
can be easily made feeble enough by taking $\ldh$ and $\ldH\sim 10^{-12}$. 
As we will see later, the other important production mechanism of $\phi_{DM}$ 
is shown by the Feynman diagram 
where $N_2$ and $N_3$ annihilate to $\phi_{DM}$ via the new gauge 
boson $\zmt$. The expression for the cross section of this process is 
given in Eq. (\ref{n1n2goingphidm}) in Appendix A. We see  
that the cross section for this process is proportional to $\sim \gmt^4\nmt^2/10^2$. 
Since we fix $\gmt = 9\times 10^{-4}$ to explain the anomalous muon $(g-2)$, 
we take $\nmt\sim 10^{-5}$ to keep 
$\sigma_{N_{j}N_{j} \rightarrow \phi_{DM}^{\dagger}\phi_{DM}}$ small 
enough so that $\phi_{DM}$ stays out of chemical equilibrium. 
This choice for $\nmt$ also ensures that there is a remnant $Z_2$ 
symmetry even when $\umt$ symmetry is broken spontaneously, which 
enables $\phi_{DM}$ to be stable. 
Thus, $\phi_{DM}$ behaves as a FIMP dark matter, it 
stays out of thermal equilibrium at all times, and 
is produced by the freeze-in mechanism.


The evolution of comoving number density of FIMP produced from the decays
as well as annihilations of the SM and BSM particles is
governed by the Boltzmann equation. The Boltzmann equation
in terms of the comoving number density of $\dm$ is given below.
This equation contains both decay as well as annihilation
terms. While deriving the Boltzmann equation for the FIMP $\dm$,
we have taken all the particles except $\dm$ in
thermal equilibrium and hence their number densities follow
the Maxwell-Boltzmann distribution function.
\begin{eqnarray}
&&\dfrac{dY_{\phi_{DM}}}{dz} = \dfrac{2 M_{pl}}{1.66 M_{h_2}^{2}}
\dfrac{z \sqrt{g_{\star}(z)}}{g_{s}(z)}\,\,\Bigg[\sum_{i = 1,\,2}
\langle\Gamma_{h_i\rightarrow } \phi_{DM}^{\dagger} \phi_{DM}\rangle
(Y_{i}^{eq} - Y_{\phi_{DM}})\Bigg]\nn \\
&&~~~~~~~~~
+\dfrac{4 \pi^{2}}{45} \dfrac{M_{pl} M_{h_2}}{1.66}
\dfrac{\sqrt{g_{\star}(z)}}
{z^{2}}\,\,\Bigg[\sum_{p = W,Z,h_{1},h_{2},f}
\langle\sigma {\rm v}_{p\bar{p}\rightarrow \phi_{DM}^{\dagger}
\phi_{DM}}\rangle (Y_{p}^{eq\,\,2} - Y_{\phi_{DM}}^{2})\nn \\
&&~~~~~~~~~+\sum_{i = 1,j = 2,3}
\langle\sigma {\rm v}_{N_{i} N_{j}\rightarrow \phi_{DM}^{\dagger}
\phi_{DM}}\rangle (Y_{N_i}^{eq} Y_{N_j}^{eq} - Y_{\phi_{DM}}^{2}) +
\langle\sigma {\rm v}_{h_{1} h_{2}\rightarrow \phi_{DM}^{\dagger}
\phi_{DM}}\rangle (Y_{h_1}^{eq} Y_{h_2}^{eq} - Y_{\phi_{DM}}^{2})
\Bigg]\,. \nn \\ 
\label{BE}
\end{eqnarray}
In the above equation $Y_{\phi_{DM}} = \dfrac{n_{\phi_{DM}}}{\rm s}$
is the comoving number density, $n_{\phi_{DM}}$
represents the actual number density of the dark matter candidate
$\phi_{DM}$ while ${\rm s}$ is the entropy of the
Universe. The quantity $z=\dfrac{\Lambda}{T}$,
where $\Lambda$ is some mass scale and here 
it corresponds to the mass of the second Higgs
$h_{2}$\,($\Lambda \sim M_{h_2}$). The temperature of the
Universe is denoted by $T$ and $M_{pl} = 1.22\times 10^{19}$ GeV
is the Planck mass. The function $g_{\star}(z)$ is related to the
degrees of freedom and has the following expression,
\begin{eqnarray}
\sqrt{g_{\star}(z)} = \dfrac{g_{\rm s}(z)}
{\sqrt{g_{\rho}(z)}}\,\left(1 -\dfrac{1}{3}
\dfrac{{\rm d}\,{\rm ln}\,g_{\rm s}(z)}{{\rm d} \,{\rm ln} z}\right)\,. 
\end{eqnarray}  
In the above equation $g_{s}(z)$ and $g_{\rho}(z)$ are
the effective degrees of freedom corresponding to
the entropy and energy densities of the Universe. 
If the decaying particles ($h_1$, $h_2$) are in thermal equilibrium
\footnote{If they are not in thermal equilibrium then we have
to find their non-thermal momentum distribution functions by solving the
appropriate Boltzmann equations.} then the thermal average of
the decay width can be expressed in a simpler form as follows,
\begin{eqnarray}
\langle\Gamma_{h_i\rightarrow } \phi_{DM}^{\dagger} \phi_{DM}\rangle\,\,
=\, \dfrac{K_{1}(z)}{K_{2}(z)}\,\,
\Gamma_{i\rightarrow } \phi_{DM}^{\dagger} \phi_{DM}\,.
\end{eqnarray}  
In the above expression $K_{1}(z)$ and $K_{2}(z)$ are
the modified Bessel's functions of order
one and two, respectively.

Next we explain the Boltzmann equation given in Eq.~(\ref{BE})
term by term. The first term on the R.H.S represents the contribution
to $Y_{\dm}$ arising from the decays of $h_1$ and $h_2$ and it is proportional
to the equilibrium comoving number density of the decaying particle. 
On the other hand, the inverse
decay term proportional to the comoving number density of $\dm$
occurs with a negative sign as it {\it washes out} the 
$\dm$ number density. However, as we have mentioned before
that the initial number density of $\dm$ is extremely small
due to its non-thermal origin and hence the negative feedback
coming from the inverse processes can be safely neglected.

Similarly, the second, third and fourth terms in the R.H.S
of Eq.~(\ref{BE}) indicate the net contribution to $Y_{\dm}$
coming from the annihilations of SM and BSM particles at
the early stage of the Universe. Unlike the first term (decay term)
of Eq.~(\ref{BE})
these terms are proportional to the second power of the
comoving number densities of relevant particles. The
term proportional to $Y^2_{p}$ ($p=W,\,Z,\,h_1,\,h_2,\,f$)
(in the second term of the Boltzmann equation)
represents the increment of $Y_{\dm}$ from the pair
annihilation of $p$ and its antiparticle while
the feedback arising from the inverse
process $\dm {\dmd} \rightarrow p \bar{p}$
is proportional to $Y^2_{\dm}$ and 
comes with a negative sign. 
The rest of the annihilation terms represent the production
processes of $\dm$ from the annihilations of
two different particles such as ($N_i,\,N_j$),
($h_1,\,h_2$) and consequently these
terms are proportional to the comoving number densities
of two different initial state particles. The terms
indicating the inverse processes ($\dm \dm^{\dagger} \rightarrow
N_i N_j$, $h_i h_j$, $i\neq j$) are proportional
to $Y^2_{\dm}$, as in the second term of the Boltzmann equation. 
All the annihilation terms of Eq.~(\ref{BE}) are proportional to the thermal
averaged annihilation cross sections of relevant processes
and if all the other particles (except
$\dm$) involved in the annihilation processes
are in thermal equilibrium then the general expression of 
thermally averaged annihilation cross section of
a process $A\,B\,\,\rightarrow \phi_{DM}^{\dagger}\, \phi_{DM}$
is given by
\begin{eqnarray}
f_{1} &=& \sqrt{s^{2} + (M_{A}^{2} - M_{B}^{2})^{2}
- 2\,s\,(M_{A}^{2} + M_{B}^{2})}\,,\nn \\
f_{2} &=& \sqrt{s - (M_{A} - M_{B})^{2}} \,\,\sqrt{s
- (M_{A} + M_{B})^{2}}\,\,, \nn \\
\langle\sigma v_{A\,B \rightarrow \phi_{DM}^{\dagger}
\phi_{DM}}\rangle &=& \dfrac{1}{8 \,M_{A}^{2}\, M_{B}^{2}\,T\,
K_{2}\left(\dfrac{M_{A}}{T}\right) \,K_{2}\left(\dfrac{M_{B}}{T}\right)}
\times \nonumber \\
&& \int_{(M_{A} + M_{B})^{2}}^{\infty}
\dfrac{\sigma_{A\,B\rightarrow} \phi_{DM}^{\dagger}
\phi_{DM}}{\sqrt{s}}\,f_{1}\,f_{2}\,K_{1}
\left(\dfrac{\sqrt{s}}{T}\right)\,{\rm d}s \,. 
\label{avg_th}
\end{eqnarray}
In the above Eq.~(\ref{avg_th}), $s$ is the Mandelstam variable,
$M_{A}$ and $M_{B}$ are the masses of the initial state particles,
$\sigma_{A\,B\rightarrow} \phi_{DM}^{\dagger} \phi_{DM}$ is the annihilation
cross section, 
while $K_{i}(x)$ is the modified Bessel function of order $i$.
In order to get the comoving number density ($Y_{\phi_{DM}}$)
of the DM particle $\phi_{DM}$, we have to solve the Boltzmann equation
numerically. After determining the comoving number
density $Y_{\phi_{DM}}$ of dark matter particle $\phi_{DM}$ at the present epoch,
we can determine the relic density \cite{Edsjo:1997bg, Biswas:2011td} from the following relation,
\begin{eqnarray}
\Omega_{\phi_{DM}} h^{2} = 2.755 \times 10^{8}\, \left(
\dfrac{M_{\phi_{DM}}}{\rm GeV}\right)\,
Y_{\phi_{DM}} (T_{0})\,,
\label{rel-den}
\end{eqnarray}
where $M_{DM}$ is in GeV. The observed value of the 
dark matter relic density given by the Planck collaboration
\cite{Ade:2015xua}, is 
\begin{eqnarray}
\Omega_{\phi_{DM}} h^{2} = 0.1197 \pm 0.0022.
\label{rel-val}
\end{eqnarray}
In this work we have used the above relic density bound.
\section{Results}
\label{res}

\begin{figure}[h!]
\centering
\includegraphics[angle=0,height=7.5cm,width=12cm]{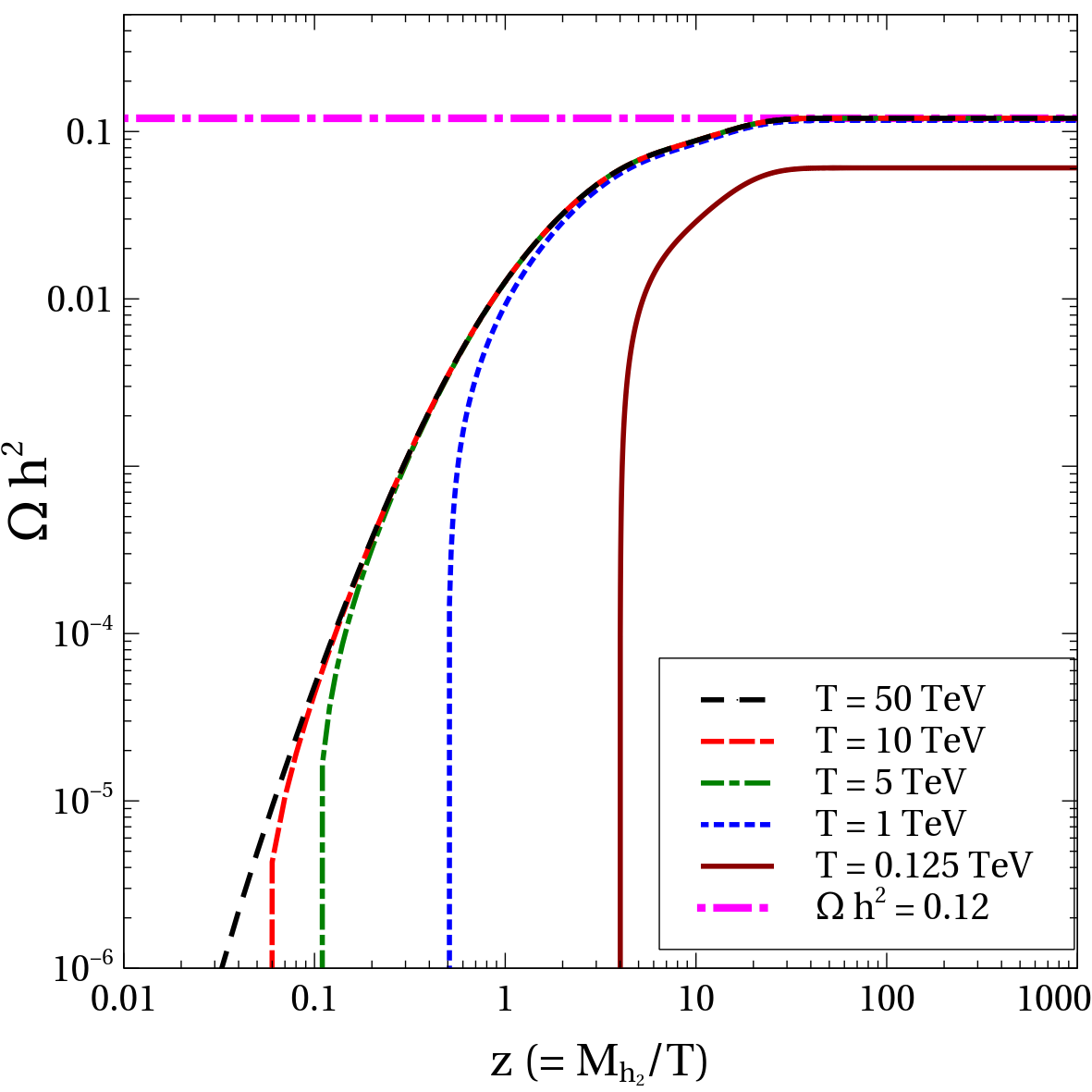}
\caption{Variation of relic density for different choices of initial
temperature while the other parameters have been kept fixed
at, $M_{Z_{\mu\tau}} = 0.1$ GeV, $g_{\mu\tau} = 9.0\times 10^{-4}$,
$\alpha = 0.01$, $\lambda_{Dh} = 9.8\times 10^{-13}$,
$\lambda_{DH} = 1.3\times 10^{-11}$, $M_{DM} = 50.0$ GeV,
$n_{\mu\tau} = 5.5\times 10^{-5}$.}
\label{1a}
\end{figure}
We have implemented our model in LanHEP \cite{Semenov:2010qt}
to generate all the vertex factors which are required
to calculate the relevant annihilation cross sections and
decay widths. Corresponding  Feynman diagrams are shown
in Fig.~\ref{feyn}. After putting all the cross sections
and decay widths (as listed in Appendix \ref{App:AppendixA}
and Appendix \ref{App:AppendixB}) in Eq.~(\ref{BE}), we 
solve the Boltzmann equation numerically and study the
related phenomenology of FIMP dark matter. Throughout this analysis
we keep the extra Higgs mass fixed at $M_{h_2} = 500$ GeV.
In Fig.~\ref{1a}, we show the variation of the dark matter relic density
with $z$ for different choices of the initial temperature.
From the figure, it is clear that if the initial
temperature is greater $T_{\rm in}\geq 1$ TeV, then there is no
dependence of the final relic density on the value of the initial temperature.
This can be explained in the following way. As the heavy Higgs
$h_2$ is in thermal equilibrium with the cosmic soup, hence
the maximum production of $\dm$ from the decay of $h_2$ occurs around
a temperature of the Universe ($T$) $\sim M_{h_2}$ i.e. 500 GeV.
However, as the temperature of the Universe drops below $M_{h_2}$,
the number density of the extra Higgs boson ($h_2$) becomes
exponentially suppressed (or Boltzmann suppressed), which
in turn reduces the final abundance of $\dm$. 
This case is shown by 
the choice $T_{\rm in} = 125\,\,\text{GeV}$ in the figure. 
Hence in what follows, 
we take a fixed $T_{\rm in}=1\,\,{\rm TeV}$.  

\begin{figure}[h!]
\centering
\includegraphics[angle=0,height=7cm,width=8cm]{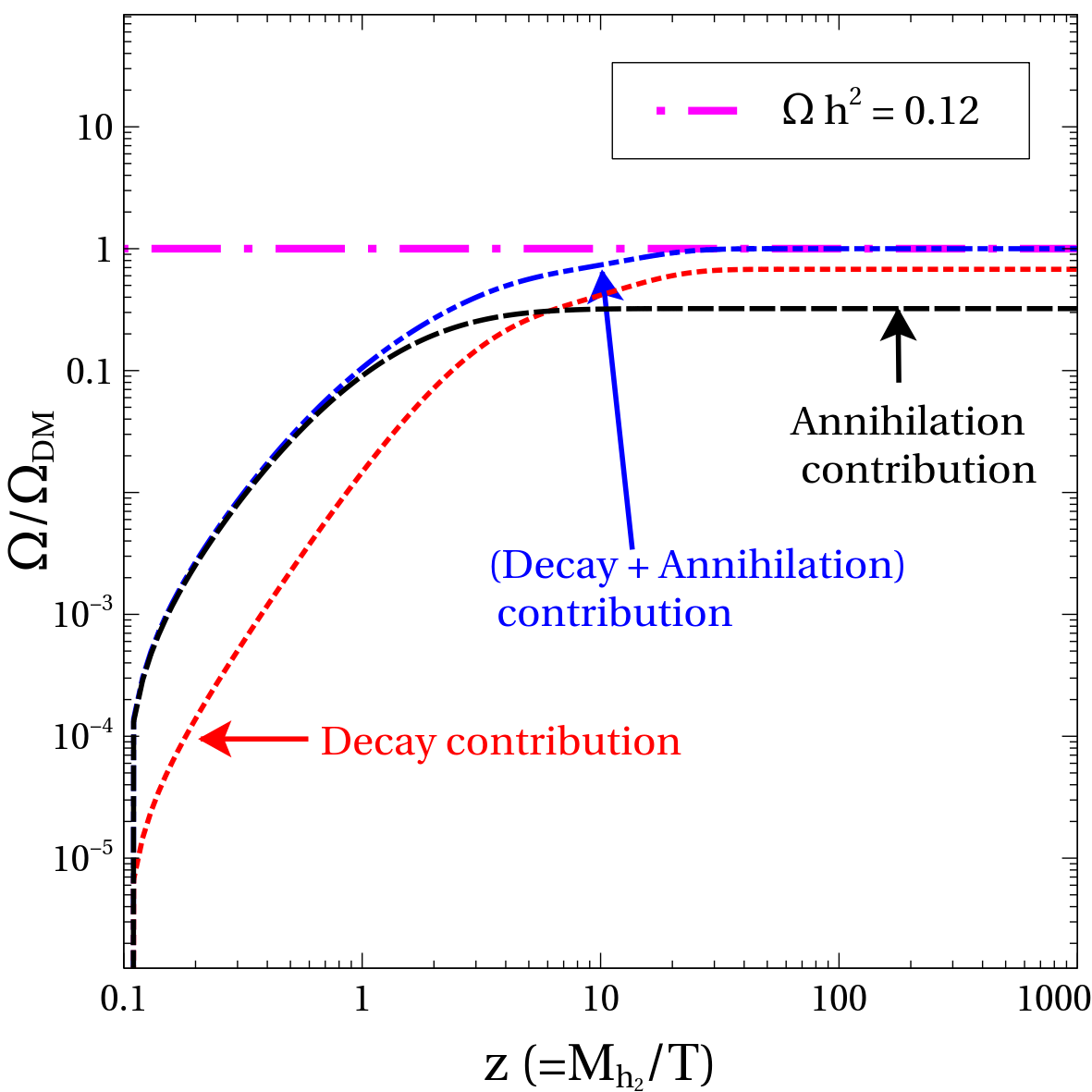}
\includegraphics[angle=0,height=7cm,width=8cm]{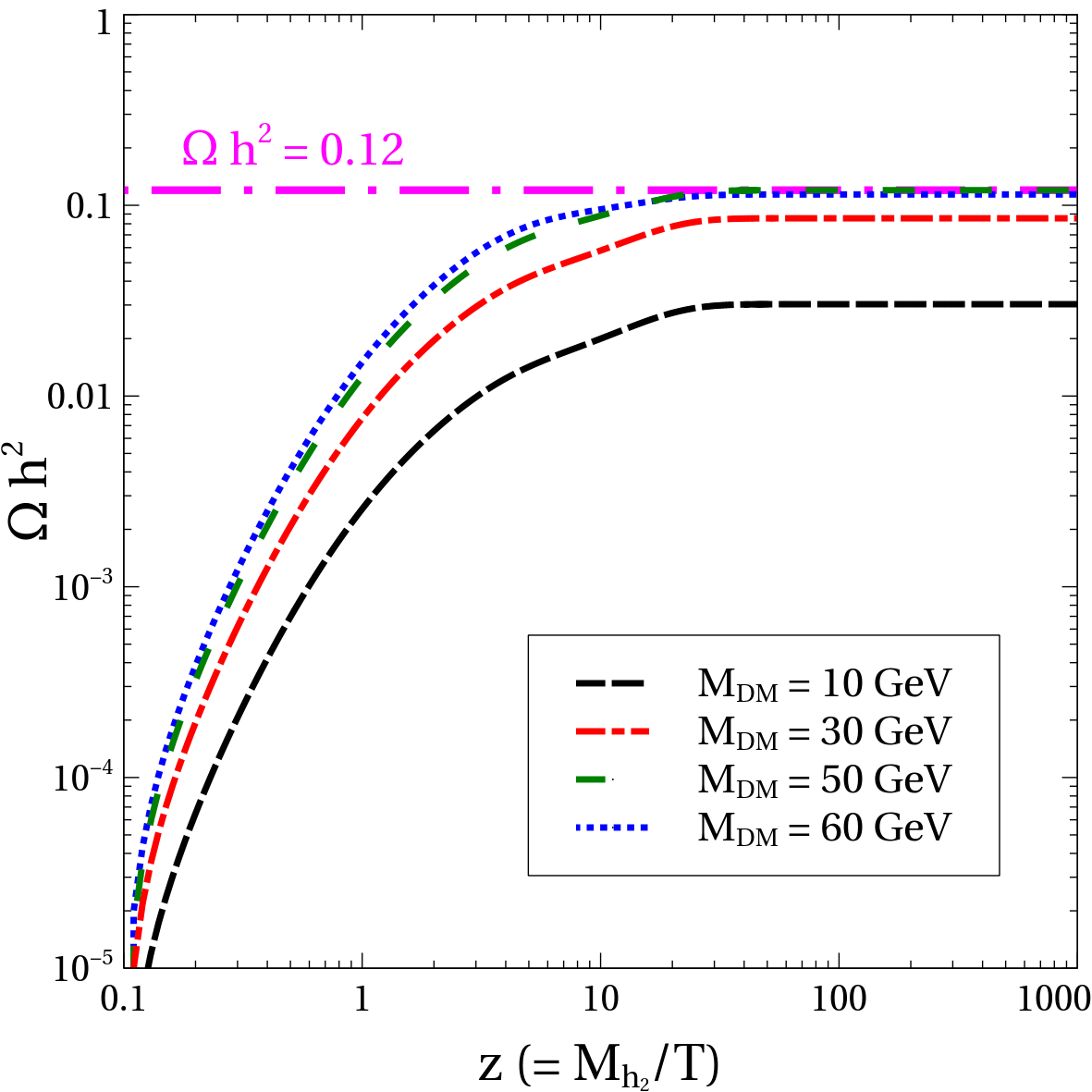}
\caption{Left panel showing the contributions of decay and
annihilation in the total relic density. Right panel:
Variation of dark matter relic density with $z$ for four different values
of dark matter mass M$_{DM}$. The other parameters are
kept fixed at $M_{Z_{\mu\tau}} = 0.1$ GeV,
$g_{\mu\tau} = 9.0\times 10^{-4}$, $\alpha = 0.01$, $\lambda_{Dh} = 9.8\times
10^{-13}$, $\lambda_{DH} = 1.3\times 10^{-11}$, $M_{DM} = 50.0$ GeV (LP),
$n_{\mu\tau} = 5.5\times 10^{-5}$.}
\label{2a}
\end{figure}

In the left panel of Fig.~\ref{2a}, we show
the relative contributions of two different types
of production processes (decay and annihilation)
to $\Omega h^2$. The red dotted line represents the
contribution from the decay of SM-like Higgs boson h$_1$ and
extra Higgs boson h$_2$, while the black dashed line corresponds to
the contribution from the all possible annihilation channels
of SM and BSM particles (see Fig.~\ref{feyn} for the
corresponding Feynman diagrams). The total contribution
towards the relic density of $\dm$ coming from the decay
as well as annihilation of different particles is
represented by blue dashed-dotted line.
The horizontal magenta line indicates the 
observed value of dark matter relic density
($\Omega h^2\sim 0.12$ \cite{Ade:2015xua})
at the present epoch. From this plot, it
can be seen clearly that for our chosen set of
model parameters ($M_{Z_{\mu\tau}} = 0.1$ GeV,
$\gmt = 9.0\times 10^{-4}$, $\alpha = 0.01$, $\lambda_{Dh} = 9.8\times
10^{-13}$, $\lambda_{DH} = 1.3\times 10^{-11}$, $M_{DM} = 50.0$ GeV,
$\nmt = 5.5\times 10^{-5}$), the decay processes contribute
$\sim 67\%$ of dark matter production while rest of the dark matter
particles are produced from the annihilations of
different SM as well as BSM particles.
In the right panel of Fig.~\ref{2a},
the variation of relic density with $z$ (i.e. with respect to
the inverse of temperature $T$) has been shown for
four different values of dark matter mass $M_{DM}$. 
For M$_{DM} =$ 10 GeV, 30 GeV and 50 GeV, the relic density
is seen to rise with M$_{DM}$. This agrees with the expression
for relic density given in Eq.~(\ref{rel-den}). 
However, if we take a slightly higher value of dark matter mass,
M$_{DM}=$ 60 GeV (blue dotted line), the relic density 
decreases instead of increasing.
This is because M$_{DM}=$ 60 GeV is very close to half of the SM
like Higgs boson mass ($\sim M_{h_1}/2$) and
the decay mode $h_1\rightarrow \dm \dm^\dagger$
becomes phase space suppressed. 
Therefore, it
reduces the contribution arising from
$h_1$ decay and hence the final relic density
of dark matter. 

\begin{figure}[h!]
\centering
\includegraphics[angle=0,height=7cm,width=8cm]
{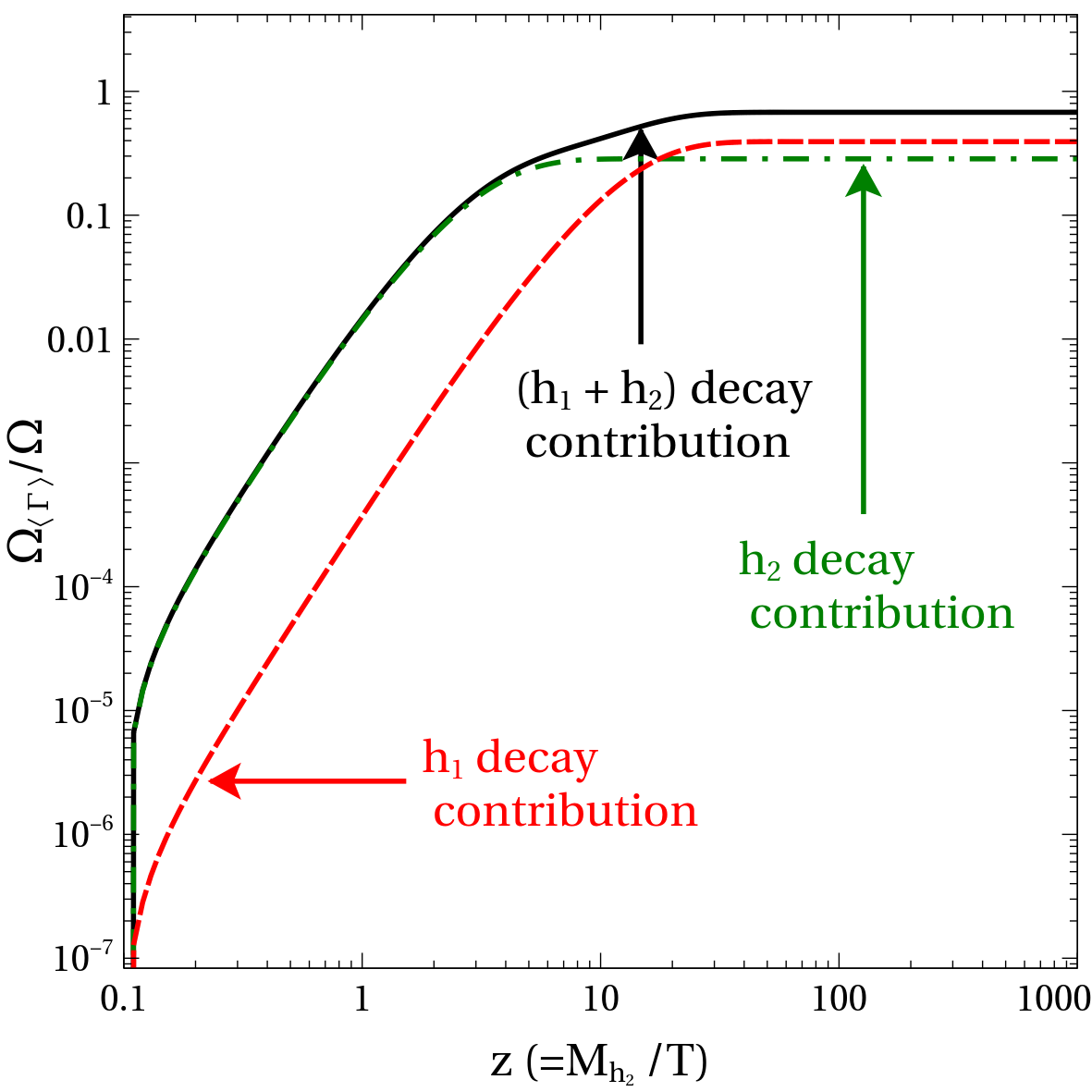}
\includegraphics[angle=0,height=7cm,width=8cm]
{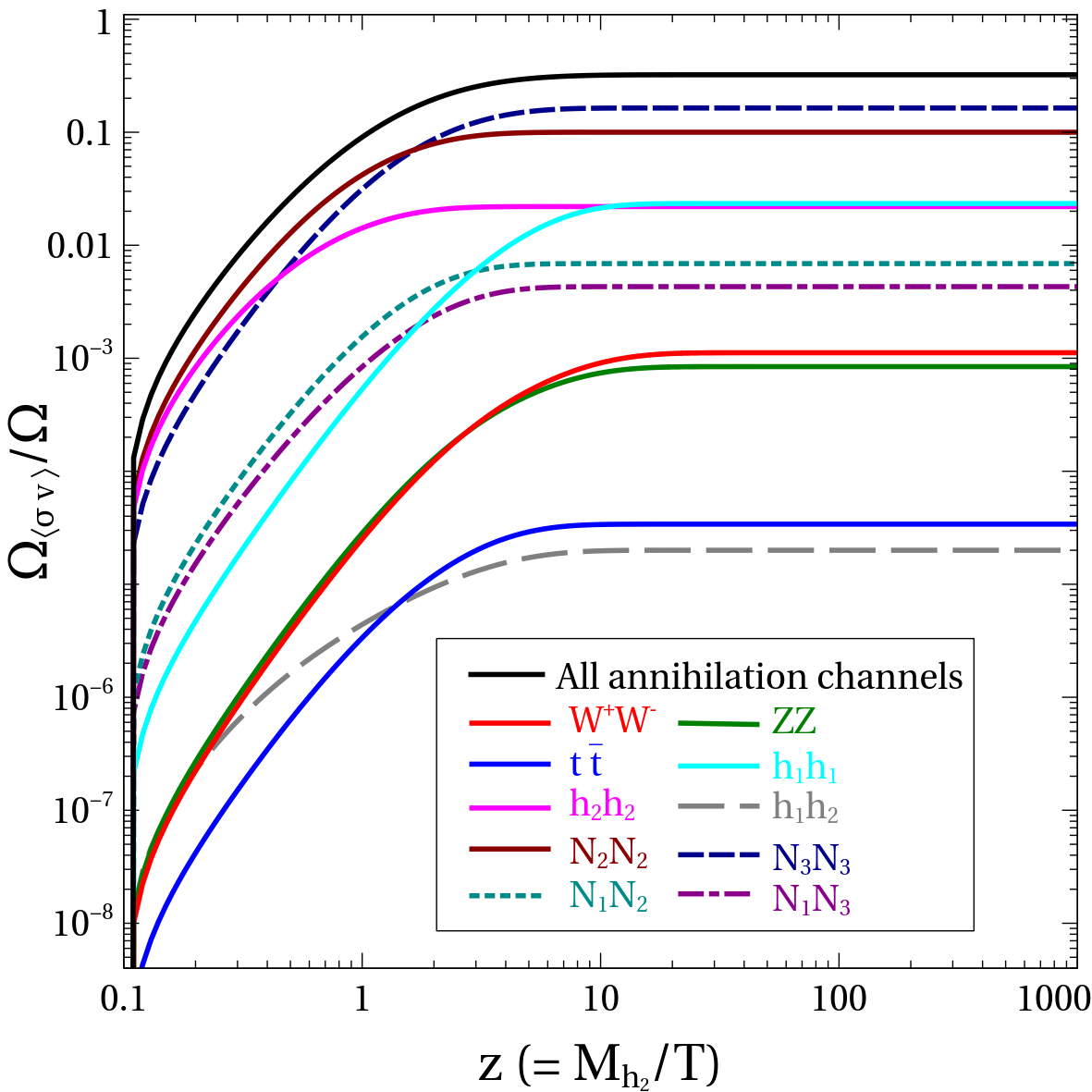}
\caption{Left Panel: Relative contributions
of two decay modes in relic density. Right Panel:
Relative contributions of different annihilation
channels towards $\Omega h^2$. Other parameters are
kept fixed at $M_{Z_{\mu\tau}} = 0.1$ GeV,
$g_{\mu\tau} = 9.0\times 10^{-4}$, $\alpha = 0.01$, $\lambda_{Dh} = 9.8\times
10^{-13}$, $\lambda_{DH} = 1.3\times 10^{-11}$, $M_{DM} = 50.0$ GeV,
$n_{\mu\tau} = 5.5\times 10^{-5}$.}
\label{3a}
\end{figure}

\begin{figure}[h!]
\centering
\includegraphics[angle=0,height=7cm,width=8cm]{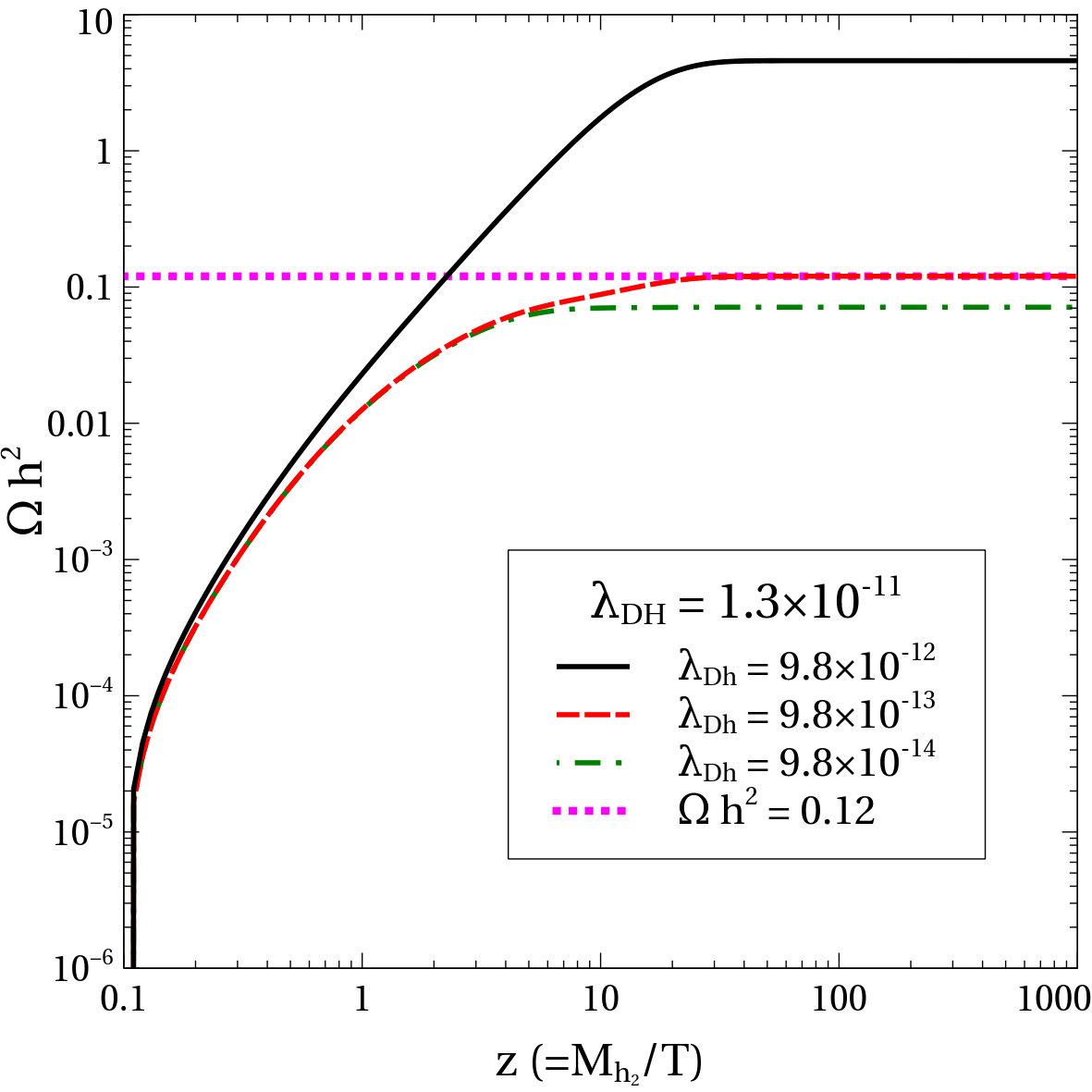}
\includegraphics[angle=0,height=7cm,width=8cm]{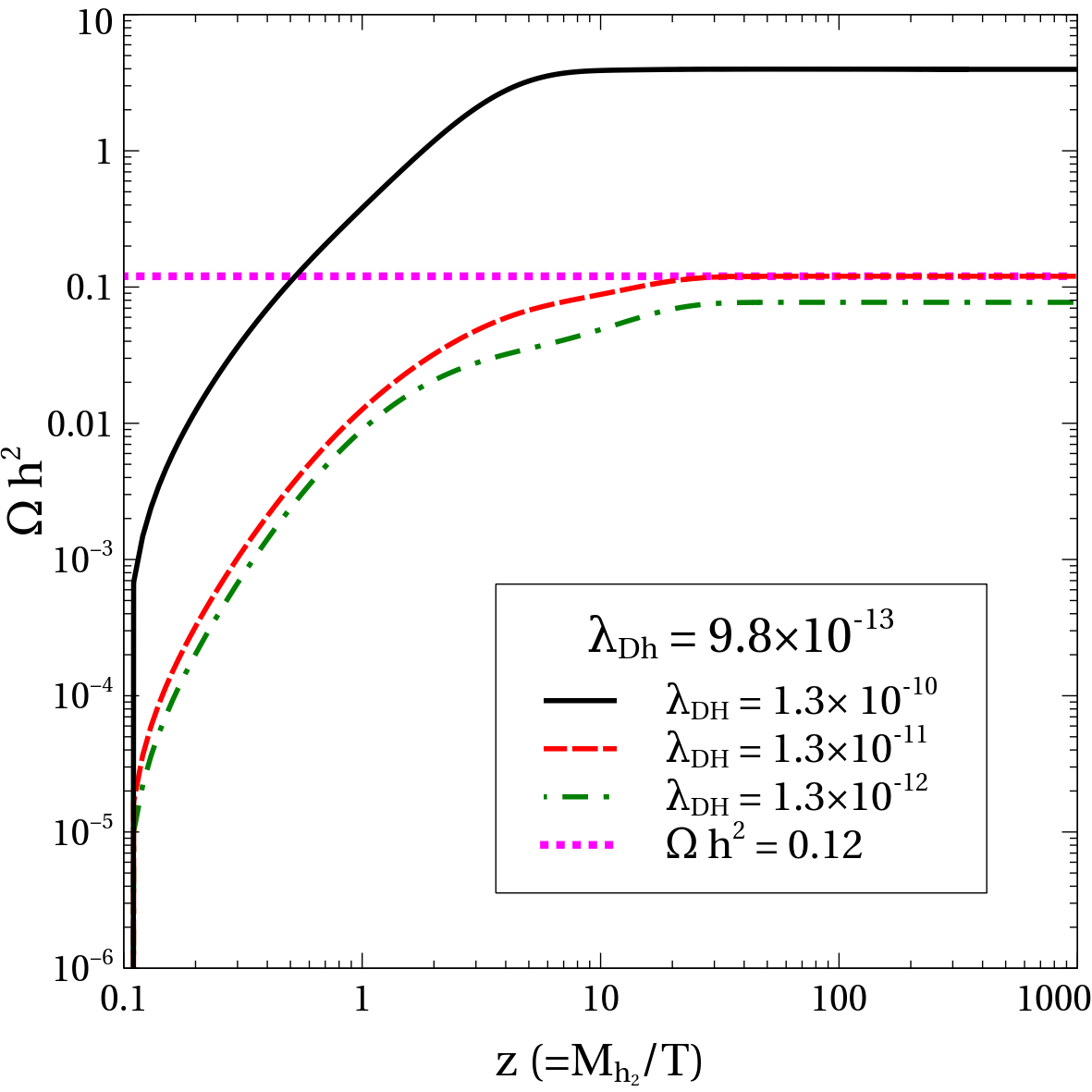}
\caption{Left (Right) Panel: Variation of relic density with $z$ for
three different values of $\lambda_{Dh}$ ($\lambda_{DH}$).
Other parameters are kept fixed at $M_{Z_{\mu\tau}} = 0.1$ GeV,
$g_{\mu\tau} = 9.0\times 10^{-4}$, $\alpha = 0.01$, $\lambda_{Dh} = 9.8\times
10^{-13}$, $\lambda_{DH} = 1.3\times 10^{-11}$, $M_{DM} = 50.0$ GeV,
$n_{\mu\tau} = 5.5\times 10^{-5}$. }
\label{4a}
\end{figure}

The contributions to $\Omega h^2$ arising from the decays of $h_1$, $h_2$
and the annihilations of SM as well
as BSM particles are shown respectively in left and right panels
of Fig.~\ref{3a}. Here we define a quantity $\frac{\Omega_{\langle
\Gamma \rangle}}{\Omega}$ ($\frac{\Omega_{\langle
\sigma {\rm v} \rangle}}{\Omega}$) which represents the
fractional contribution of a particular decay (annihilation)
channel to dark matter relic density. In the left-panel of Fig.~\ref{3a},
the contribution from h$_2$ decay has been
shown by the green dashed-dotted line and
that from h$_1$ decay has been shown by the
red dashed line, while the total decay contribution
to the dark matter relic density is represented by the black
solid line. From this plot one
can see that, initially for low values of $z$
($z<10$, corresponding to higher temperatures), 
the extra Higgs contribution to $\Omega h^2$ is more
because of its high mass. On the other hand, for higher
values of $z$ ($z>10$), the SM-like Higgs decay contribution 
starts dominating. In the right panel of Fig.\,{\ref{3a}},
we show the contribution coming 
from different annihilation channels.
The total contribution from all the annihilation
channels is represented by the black solid line while
the other lines show the contribution of individual channels.
From this plot it is clearly seen that, the two dominating
annihilation channels are $N_2\,N_2$ and
$N_3\,N_3$. Annihilation channels $h_1\,h_1$ and
$h_2\,h_2$ also have significant role in the production
processes of dark matter, while the effect of other channels are
sub dominant.     
In the left-panel of Fig.~\ref{4a}, variation of relic density with $z$ for three
different value of $\lambda_{Dh}$ have been shown. The red dashed line 
for $\lambda_{Dh} = 9.8\times 10^{-13}$ gives the correct relic density.  
In the right-panel of Fig.\,{\ref{4a}} we show the variation of
relic density for different values of the other quartic
coupling $\lambda_{DH}$. 
It is clear from Fig.~\ref{4a} that the
relic density increases with both $\lambda_{Dh}$ and
$\lambda_{DH}$ as the production modes of $\dm$
are proportional to these quartic couplings. However,
the increment of $\Omega h^2$ with respect to
increasing $\lambda_{Dh}$ or $\lambda_{DH}$
is not uniform. When we decrease $\ldh$ ($\ldH$)
from $9.8\times 10^{-13}$ ($1.3\times 10^{-11}$)
by one order of magnitude, the decrease in $\Omega h^2$ 
is very small because in this regime we have dominant
contribution from the $\zmt$ mediated right-handed neutrino
annihilation channel. However, if we increase 
$\ldh$ ($\ldH$) from $9.8\times 10^{-13}$ ($1.3\times 10^{-11}$)
by one order of magnitude, $\Omega h^2$ increases by 
more than order of magnitude since in this case the
contribution from decay channels become dominant.

\begin{figure}[h!]
\centering
\includegraphics[angle=0,height=7cm,width=8cm]{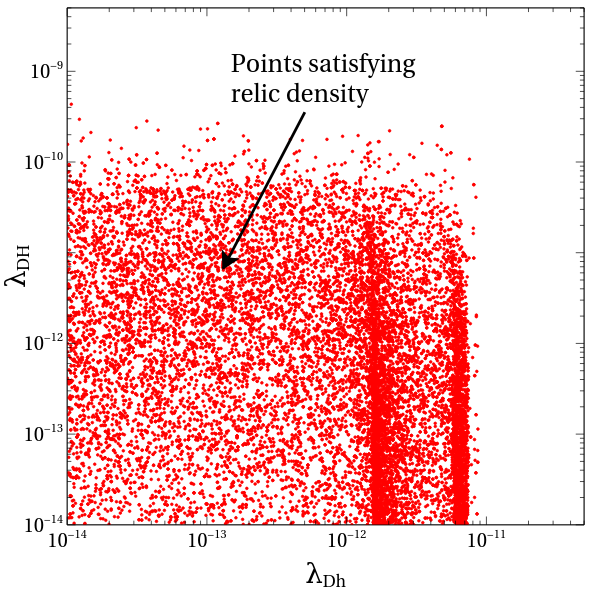}
\includegraphics[angle=0,height=7cm,width=8cm]{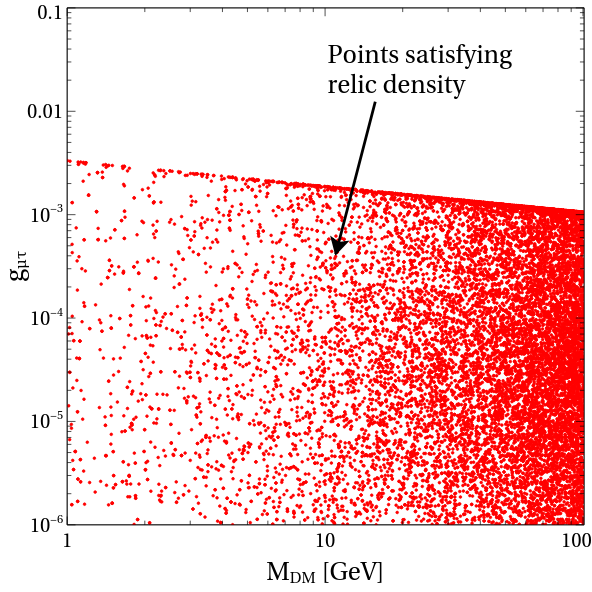}
\caption{Left panel: Allowed parameter space in
$\lambda_{Dh}$ - $\lambda_{DH}$ plane. Right panel:
Allowed parameter space in $M_{DM}$ - $g_{\mu\tau}$ plane.
In both the plots red points satisfy the relic density bound.}
\label{5a}
\end{figure}

In the left-panel of Fig.~\ref{5a}, we have shown the allowed regions in the
$\lambda_{Dh}$ - $\lambda_{DH}$ plane. The red points in the plane 
satisfy the relic density bound. Both the parameters $\lambda_{Dh}$ and 
$\lambda_{DH}$ have been varied from 10$^{-14}$ to 10$^{-8}$.
We see from the figure that for $\lambda_{Dh} \geq 8
\times 10^{-11}$ and $\lambda_{Dh} \geq 3 \times 10^{-10}$
no red points exist, and therefore these regions are disallowed by the relic
density bound. One can also notice that there is no lower bound on
$\lambda_{Dh}$ and $\lambda_{DH}$. This is because for lower values of 
$\lambda_{Dh}$ and $\lambda_{DH}$, even though 
the Higgs mediated annihilation and decay contributions become very less, 
the $Z_{\mu\tau}$ mediated annihilation channels
($N_{i}\,{N_{i}}\rightarrow \phi_{DM}^{\dagger} \,\phi_{DM}$, $i = 2,3$
see Appendix \ref{App:AppendixA}) contribute fully and hence can explain the 
relic density bound. 
In the right-panel of Fig.~\ref{5a}, we show the allowed regions in the 
$M_{DM}$ - $g_{\mu\tau}$ plane. Here we have varied dark matter 
mass $M_{DM}$ from 1 GeV to
100 GeV and the $\umt$ gauge coupling $g_{\mu\tau}$
from 10$^{-6}$ to 0.1. The figure shows that the whole
range of dark matter mass $M_{DM}$ can satisfy the relic density
bound. However, the gauge coupling $g_{\mu\tau} \geq 3
\times 10^{-3}$ does not satisfy the relic density
bound as over production of $\dm$ occurs
through the annihilation channel
$N_{i}\,{N_{i}}\rightarrow \phi_{DM}^{\dagger}
\,\phi_{DM}$, $i = 2,3$.   

\begin{figure}[h!]
\centering
\includegraphics[angle=0,height=7cm,width=8cm]{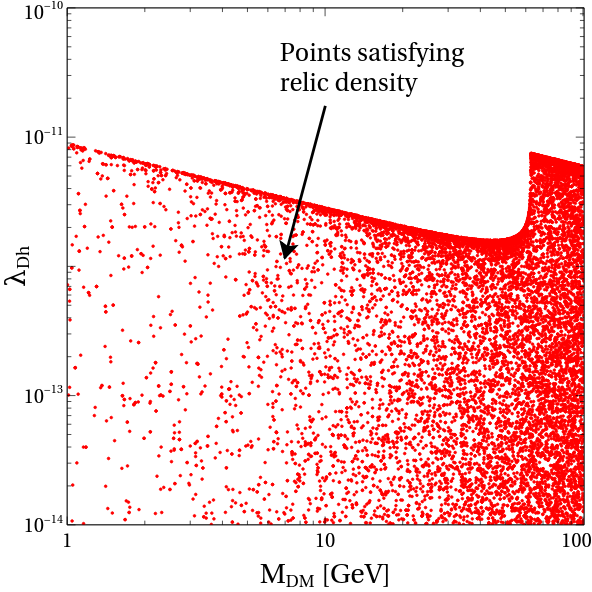}
\includegraphics[angle=0,height=7cm,width=8cm]{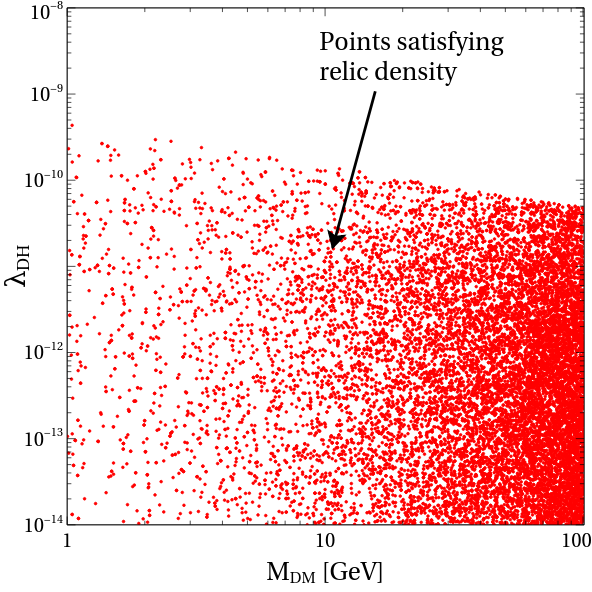}
\caption{Left panel: Allowed parameter space in $M_{DM}$ - $\lambda_{Dh}$
plane. Right panel: Allowed parameter space in $M_{DM}$ - $\lambda_{DH}$ space.
In both the plots red points satisfy the relic density bound.}
\label{6a}
\end{figure}
In Fig.~\ref{6a}, we show the allowed parameter space in the $M_{DM}-\lambda_{Dh}$
and $M_{DM}-\lambda_{DH}$ planes in the left and right panels respectively. 
As it was seen earlier, here too the whole range of considered
dark matter mass \footnote{In all the plots the allowed red points
in the $M_{DM}$ appear more dense on the right since 
we have generated the random number in linear scale
but plotted the figures in the log scale.}
is allowed. However, there is an upper limit on both 
$\lambda_{Dh}$ and $\lambda_{DH}$. In both the panels,
there exist an anti-correlation between the dark matter mass and
the quartic coupling $\lambda_{Dh(H)}$ since the 
relic density is proportional to the dark matter mass $M_{DM}$
as well as the coupling constant. Hence, if we increase $M_{DM}$
then to satisfy the relic density, $\lambda_{Dh(H)}$ must decrease. 
In the left-panel we observe that around $M_{DM}\sim 62$ GeV,
there is a rise in $\lambda_{Dh}$. This happens because
this is the resonance region for the SM-like Higgs and as a result 
there is little contribution from the decay
of SM-like Higgs (phase space suppression). Hence,
to satisfy the relic density bound, there is a sudden 
rise in $\lambda_{Dh}$ to increase the contribution
arising from the decay as well as annihilation
processes involving $h_1$. Beyond $M_{DM}\sim62.5$ GeV,
there is no contribution from the SM-like Higgs. 
Hence, the coupling $\lambda_{Dh}$ again starts
behaving in the normal way (anti-correlation). 
No such peculiar behaviour is seen for
$\lambda_{DH}$ because this parameter is important for
the decay of $h_{2}$ and the chosen mass range of the
dark matter is not in the resonance region of the extra Higgs $h_{2}$
($M_{h_2}\sim$ 500 GeV). Therefore, for the quartic coupling
$\lambda_{DH}$, the anti-correlation exists for the entire
range of dark matter mass $M_{DM}$ (1-100 GeV).    
\section{Conclusion}
\label{con}

In this work, we propose a framework for non-thermal production
of a FIMP type dark matter in the gauged U(1)$_{L_{\mu} - L_{\tau}}$ extension
of the SM. The particle content of our model includes two 
SM gauge singlet scalars and three right-handed neutrinos. 
Both the SM singlet scalars carry U(1)$_{L_{\mu} - L_{\tau}}$ charge.
One of them picks up a VEV, breaking U(1)$_{L_{\mu} - L_{\tau}}$ 
spontaneously, thereby giving the new gauge boson $\zmt$ mass. 
The new boson $\zmt$ gives additional contribution to muon $(g-2)$ 
which can explain its measured value. The U(1)$_{L_{\mu} - L_{\tau}}$ 
breaking also leads to additional terms in the light neutrino mass matrix 
which has its origin via the Type-I seesaw mechanism. This model can 
therefore explain consistently the neutrino masses and mixing patterns 
observed in neutrino oscillation experiments. Finally, the other SM singlet 
scalar $\dm$ does not acquire any VEV and remains stable due to a remnant 
$Z_2$ symmetry even after the breaking of U(1)$_{L_{\mu} - L_{\tau}}$ 
due to our choice of its $L_{\mu}-L_{\tau}$ charge ($\nmt$). 
This scalar can therefore serve as a dark matter candidate. 

In this model,
$\dm$ can interact with the visible sector (including both
SM as well as BSM particles) only through the scalar bosons $h_1$,
$h_2$ and $L_{\mu}-L_{\tau}$ gauge boson $\zmt$. Hence, in
order to keep $\dm$ out of thermal equilibrium we have
considered the corresponding couplings of $\dm$ to be extremely
feeble. All other particles (both SM and BSM) in the early Universe,
however remain in equilibrium with the thermal soup.
We have numerically solved the Boltzmann equation
containing all possible production modes of $\dm$ from
decays and annihilations of SM and BSM particles. We have found
that, among all of its production modes,  $\dm$ is mainly produced
from the decays of $h_1$, $h_2$ and the pair annihilations of right
handed neutrinos mediated by $\zmt$. The latter process beautifully 
sets the interplay between the dark matter sector, neutrino masses 
and muon $(g-2)$.
The solution of Boltzmann
equation also indicates that, for our chosen set of model
parameters, decay processes of $h_1$ and $h_2$ contribute
$\sim 67\%$ of the total dark matter relic abundance while rest $\sim 33\%$
contribution is coming from the annihilation of particles (mainly
right-handed neutrinos) in the thermal bath. We have shown that,
the relic abundance of our FIMP type dark matter candidate $\dm$ lies
within the Planck Limit (0.1172 $\leq \Omega h^2 \leq 0.1224$) only when
the relevant parameters satisfy the following bounds:
$\ldh \la 10^{-11}$, $\ldH \la 10^{-10}$ and $\gmt \leq 3 \times 10^{-3}$
for the dark matter mass range of 1 GeV to 100 GeV and
$\nmt = 5.5 \times 10^{-5}$. 
Consequently, due to the low coupling
strength there does not exist any bound on dark matter spin independent
scattering cross section from direct detection experiments.


In conclusion, our proposed U(1)$_{L_{\mu} - L_{\tau}}$ extension of the SM 
can explain the three main puzzles that demand beyond Standard Model physics. 
First, it can successfully explain the smallness of neutrino masses via the Type-I 
seesaw mechanism as well as the peculiar mixing pattern of the neutrinos 
via the U(1)$_{L_{\mu} - L_{\tau}}$ gauge symmetry that also 
acts on the lepton flavours, thereby giving a pattern to the
light neutrino mass matrix. 
Second, the additional one loop contribution of the extra neutral gauge boson 
$Z_{\mu\tau}$ can successfully satisfy the muon ($g-2$) data. 
And finally, the model has a SM singlet scalar with non-zero U(1)$_{L_{\mu} - L_{\tau}}$,
that makes it stable and  
which acts as a non-thermal dark matter candidate, thereby satisfying
constraint on the relic abundance and at the same time evading
all bounds coming from direct and indirect dark matter
detection experiments. 

\section{Acknowledgements} 
The authors would like to thank the Department of Atomic Energy
(DAE) Neutrino Project under the XII plan of Harish-Chandra
Research Institute. SK and AB also acknowledge the cluster
computing facility at HRI (http://cluster.hri.res.in).
This project has received funding from the European Union's Horizon
2020 research and innovation programme InvisiblesPlus RISE
under the Marie Sklodowska-Curie
grant  agreement  No  690575. This  project  has
received  funding  from  the  European
Union's Horizon  2020  research  and  innovation
programme  Elusives  ITN  under  the 
Marie  Sklodowska-
Curie grant agreement No 674896.

\appendix
\section*{Appendix}
\section{Analytical Expression of relevant Cross sections}
\label{App:AppendixA}
In this section we have given all the relevant cross sections which are required
to solve the Boltzmann equation (Eq.~(\ref{BE})) numerically.
We have expressed all the cross sections in terms of three
Mandelstam variables. Here, we have used the
usual particle notation as we have given in the model section.
The vertex factors which are common to the most of the
annihilation diagrams are,

\begin{eqnarray}
g_{h_{1} \phi_{DM} \phi_{DM}} &=&
- \left(v\lambda_{Dh}\cos \alpha +
v_{\mu\tau}\lambda_{DH}\sin \alpha\right)\,, \nn \\
g_{h_{2} \phi_{DM} \phi_{DM}} &=&
\left(v\lambda_{Dh}\sin \alpha -
v_{\mu\tau}\lambda_{DH}\cos \alpha\right)\,.
\label{dm_vertex}
\end{eqnarray}
\\
\begin{itemize}
\item \underline{$W^{+}$\,$W^{-}$ $\rightarrow
\phi_{DM}^{\dagger}\,\phi_{DM}$}\,:
\newline
\newline
The annihilation of the vector bosons $W^{+}$ and $W^{-}$ into the dark matter
$\phi_{DM}$ and $\phi_{DM}^{\dagger}$ are occurred by the two s channel diagrams
which are mediated by the SM-like Higgs $h_1$ and the SM extra Higgs boson $h_2$.
Expression of this annihilation cross section is,
\begin{eqnarray}
g_{h_{1}WW} &=& \dfrac{2M_{W}^{2} \cos\alpha}{v}\,,\nn \\
g_{h_{2}WW} &=& -\dfrac{2M_{W}^{2} \sin\alpha}{v}, \nn \\
A_{WW} &=& \dfrac{g_{h_{1}WW}\,\,g_{h_{1}\phi_{DM}^{\dagger}
\phi_{DM}}}{(s-M_{h_1}^{2}) + i M_{h_1} \Gamma_{h_1}}
+ \dfrac{g_{h_{2}WW}\,g_{h_{2}\phi_{DM}^{\dagger}\phi_{DM}}}
{(s-M_{h_2}^{2}) + i M_{h_2} \Gamma_{h_2}}, \nn \\
M_{WW} &=& \dfrac{2}{9}\,\left(1 + \dfrac{(s - 2M_{W}^{2})^{2}}
{8M_{W}^{4}}\right)\,A_{WW}, \nn \\
\sigma_{WW \rightarrow \phi_{DM}^{\dagger}\phi_{DM}} &=& \dfrac{1}{16 \pi s}\,\,
\sqrt{\dfrac{s - 4M_{DM}^{2}}{s - 4M_{W}^{2}}} \,\,\,|M_{WW}|^{2}\,.
\end{eqnarray}
where $|M_{WW}|^{2}$ represents the square of the absolute value of $M_{WW}$.
The couplings $g_{h_{1}\phi_{DM}^{\dagger}\phi_{DM}}$ and
$g_{h_{2}\phi_{DM}^{\dagger}\phi_{DM}}$
are given in Eq.~(\ref{dm_vertex}) while $\Gamma_{h_1}$ and
$\Gamma_{h_2}$ are the total decay widths of the SM-like
Higgs and the extra Higgs respectively (see Appendix \ref{App:AppendixB}
for more details). In the expression of $M_{WW}$
the extra factor 1/9 comes due to average of the polarisation vectors
of two initial state $W$ bosons. 

\item \underline{$Z\,Z \rightarrow
\phi_{DM}^{\dagger}\,\phi_{DM}$}\,:
\newline
\newline
The self annihilation of $Z$ boson to dark matter
particles $\phi_{DM}$ and $\phi_{DM}^{\dagger}$ has two $s$\,channel
diagrams, one of them is mediated by the SM-like Higgs $h_1$ while
another one is mediated by the extra Higgs $h_2$. Cross section structure
for the $Z\,Z$ annihilation is similar to $W^{+}\,W^{-}$ annihilation
and has the following form,
\begin{eqnarray}
g_{h_{1}ZZ} &=& \dfrac{2M_{Z}^{2} \cos \alpha}{v}\,,\nn \\
g_{h_{2}ZZ} &=& -\dfrac{2M_{Z}^{2} \sin \alpha}{v}, \nn \\
A_{ZZ} &=& \dfrac{g_{h_{1}ZZ}\,g_{h_{1}\phi_{DM}^{\dagger}
\phi_{DM}}}{(s-M_{h_1}^{2}) + i M_{h_1} \Gamma_{h_1}}
+ \dfrac{g_{h_{2}ZZ}\,g_{h_{2}\phi_{DM}^{\dagger}\phi_{DM}}}
{(s-M_{h_2}^{2}) + i M_{h_2} \Gamma_{h_2}}, \nn \\
M_{ZZ} &=& \dfrac{2}{9}\,\,\left(1 + \dfrac{(s - 2M_{Z}^{2})^{2}}
{8M_{Z}^{4}}\right)\,A_{ZZ}, \nn \\
\sigma_{ZZ \rightarrow \phi_{DM}^{\dagger}\phi_{DM}}
&=& \dfrac{1}{16 \pi s}\,\,\sqrt{\dfrac{s - 4M_{DM}^{2}}
{s - 4M_{Z}^{2}}} \,\,\,|M_{ZZ}|^{2}\,.
\end{eqnarray}
\item \underline{$h_{1}$\,$h_{1}$ $\rightarrow
\phi_{DM}^{\dagger}\,\phi_{DM}$}\,:
\newline
\newline
The annihilation of two $h_1$ to $\phi_{DM}^{\dagger}\,\phi_{DM}$
occurs through one four points interaction process
as well as two $s$\,channel processes \footnote{While calculating
$h_i\,h_j\rightarrow \dm \dm^{\dagger}$,
we have neglected the subdominant $t$\,channel
interaction process mediated by $\dm$.}
mediated by the SM-like Higgs $h_1$ and extra Higgs $h_2$.
The expression of the cross section is the following,
\begin{eqnarray}
&&g_{h_{1}h_{1}h_{1}} = -3\,[2\,v\lambda_{h_1}\cos^{3}\alpha
+ 2\,v_{\mu\tau}\,\lambda_{h_2}\sin^{3}\alpha +
\lambda_{h_{1}h_{2}}\sin\alpha\,\cos\alpha\,
(v\sin\alpha + v_{\mu\tau}\cos\alpha)],\nn \\
&&g_{h_{1}h_{1}h_{2}} = [6\,v\lambda_{h_1}\cos^{2}\alpha\sin\alpha -
6\,v_{\mu\tau}\lambda_{h_2}\sin^{2}\alpha\,\cos\alpha 
-(2-3\,\sin^{2}\alpha)\,v\,\lambda_{h_{1}h_{2}}\,\sin\alpha
\nn\\ 
&&~~~~~~~~~-(1-3\sin^{2}\alpha)v_{\mu\tau}\,
\lambda_{h_{1}h_{2}}\cos\alpha] \,,\label{h2h1h1} \\
&&g_{h_{1}h_{1}\phi_{DM}^{\dagger}\phi_{DM}} =
-(\lambda_{Dh} \cos^{2}\alpha + \lambda_{DH} \sin^{2}\alpha) \,,\nn \\
&&M_{h_{1} h_{1}} = \dfrac{g_{h_{1}h_{1}h_{1}}\,
\,g_{h_{1}\phi_{DM}^{\dagger}\phi_{DM}}}{(s-M_{h_1}^{2}) +
i M_{h_1} \Gamma_{h_1}} + \dfrac{g_{h_{1}h_{1}h_{2}}
\,\,g_{h_{2}\phi_{DM}^{\dagger}\phi_{DM}}}{(s-M_{h_2}^{2}) +
i M_{h_2} \Gamma_{h_2}} - g_{h_{1}h_{1}\phi_{DM}^{\dagger}\phi_{DM}}, \nn \\
&&\sigma_{h_{1}h_{1} \rightarrow \phi_{DM}^{\dagger}\phi_{DM}} =
\dfrac{1}{16 \pi s}\,\,\sqrt{\dfrac{s - 4M_{DM}^{2}}{s - 4M_{h_1}^{2}}}
\,\,\,|M_{h_{1}h_{1}}|^{2}\,.
\end{eqnarray}
\item \underline{$h_{2}$\,$h_{2}$ $\rightarrow
\phi_{DM}^{\dagger}\,\phi_{DM}$}\,:
\newline
\newline
Similarly, the annihilation of two $h_2$ to $\phi_{DM}^{\dagger}\,\phi_{DM}$
also occurs through one four points interaction process as
well as two $s$\,channel processes mediated by the SM
like Higgs $h_1$ and extra Higgs $h_2$.
The expression of annihilation cross section for
$h_{2}$\,$h_{2}$ $\rightarrow \phi_{DM}^{\dagger}\,\phi_{DM}$ is
given by
\begin{eqnarray}
&&g_{h_{2}h_{2}h_{2}} = 3\,[2\,v\lambda_{h_1}\sin^{3}
\alpha - 2\,v_{\mu\tau}\lambda_{h_2}\cos^{3}\alpha +
\lambda_{h_{1}h_{2}}\sin\alpha\cos\alpha\,
(v\cos\alpha - v_{\mu\tau}\sin\alpha)],\nn \\
&&g_{h_{2}h_{2}h_{1}} = -[6\,v\lambda_{h_1}\sin^{2}
\alpha\cos\alpha + 6\,v_{\mu\tau}\lambda_{h_2}\cos^{2}\alpha\sin\alpha 
-(2-3\,\sin^{2}\alpha)v_{\mu\tau}\lambda_{h_{1}h_{2}}\sin\alpha
\,\,\nn \\
&&~~~~~~~~~+(1-3\sin^{2}\alpha)v\lambda_{h_{1}h_{2}}\cos\alpha]\,, \nn \\
&&g_{h_{2}h_{2}\phi_{DM}^{\dagger}\phi_{DM}} = -(\lambda_{Dh} \sin^{2}\alpha + \lambda_{DH} \cos^{2}\alpha) \,,\nn \\
&&M_{h_{2}h_{2}} = \dfrac{g_{h_{2}h_{2}h_{1}}\,
\,g_{h_{1}\phi_{DM}^{\dagger}\phi_{DM}}}{(s-M_{h_1}^{2}) +
i M_{h_1} \Gamma_{h_1}} + \dfrac{g_{h_{2}h_{2}h_{2}}
\,\,g_{h_{2}\phi_{DM}^{\dagger}\phi_{DM}}}{(s-M_{h_2}^{2}) +
i M_{h_2} \Gamma_{h_2}} - g_{h_{2}h_{2}\phi_{DM}^{\dagger}\phi_{DM}}, \nn \\
&&\sigma_{h_{2}h_{2} \rightarrow \phi_{DM}^{\dagger}\phi_{DM}} =
\dfrac{1}{16 \pi s}\,\,\sqrt{\dfrac{s - 4M_{DM}^{2}}{s - 4M_{h_2}^{2}}}
\,\,\,|M_{h_{2}h_{2}}|^{2}\,.
\end{eqnarray}
\\
\item \underline{$h_{1}$\,$h_{2}$ $\rightarrow
\phi_{DM}^{\dagger}\,\phi_{DM}$}\,:
\newline
\newline
Like the previous two cases, here we also have three processes
(one four points interaction and two $s$ channels) contributing to
the annihilation of $h_{1}$\,$h_{2}$ $\rightarrow
\phi_{DM}^{\dagger}\,\phi_{DM}$. The expression
of $h_{1}$\,$h_{2}$ $\rightarrow
\phi_{DM}^{\dagger}\,\phi_{DM}$ has the following
form
\begin{eqnarray}
&&g_{h_{1}h_{1}h_{2}} = [6\,v\lambda_{h_1}\cos^{2}\alpha\,\sin\alpha
- 6\,v_{\mu\tau}\,\lambda_{h_2}\sin^{2}\alpha\,\cos\alpha 
-(2-3\,\sin^{2}\alpha)\,v\,\lambda_{h_{1}h_{2}}\sin\alpha \nn \\
&&~~~~~~~~~
-(1-3\sin^{2}\alpha)v_{\mu\tau}\,\lambda_{h_{1}h_{2}}\cos\alpha] \nn \\
&&g_{h_{2}h_{2}h_{1}} = -[6\,v\lambda_{h_1}\sin^{2}\alpha\,\cos\alpha
+ 6\,v_{\mu\tau}\lambda_{h_2}\cos^{2}\alpha\sin\alpha 
-(2-3\,\sin^{2}\alpha)\,v_{\mu\tau}\,\lambda_{h_{1}h_{2}}\sin\alpha \nn \\
&&~~~~~~~~~ 
+(1-3\sin^{2}\alpha)v\lambda_{h_{1}h_{2}}\cos\alpha]\,, \nn \\
&&g_{h_{1}h_{2}\phi_{DM}^{\dagger}\phi_{DM}} = \sin\alpha \cos\alpha(\lambda_{Dh} -
\lambda_{DH}) \,,\nn \\
&&M_{h_{1}h_{2}} = \dfrac{g_{h_{1}h_{1}h_{2}}\,
\,g_{h_{1}\phi_{DM}^{\dagger}\phi_{DM}}}{(s-M_{h_1}^{2}) +
i M_{h_1} \Gamma_{h_1}} + \dfrac{g_{h_{2}h_{2}h_{1}}
\,\,g_{h_{2}\phi_{DM}^{\dagger}\phi_{DM}}}{(s-M_{h_2}^{2}) +
i M_{h_2} \Gamma_{h_2}} - g_{h_{1}h_{2}\phi_{DM}^{\dagger}\phi_{DM}}, \nn \\
&&\sigma_{h_{1}h_{2} \rightarrow \phi_{DM}^{\dagger}\phi_{DM}} =
\dfrac{1}{16 \pi s}\,\,\sqrt{\dfrac{s(s - 4M_{DM}^{2})}
{(s-(M_{h_1}+M_{h_2})^{2})(s - (M_{h_2} - M_{h_1})^{2})}}
\,\,\,|M_{h_{1}h_{2}}|^{2}\,.
\end{eqnarray}
\item \underline{$t\bar{t}$\, $\rightarrow
\phi_{DM}^{\dagger}\,\phi_{DM}$}\,:
\newline
\newline
Here, the annihilating particles are $t$,\,\,$\bar{t}$ and the final particles
are $\phi_{DM}$,\,\,$\phi_{DM}^{\dagger}$. This annihilation is occurred
only by the two $s$\,channel diagrams mediated via
$h_1$ and $h_2$ respectively. The expression of the cross section
for this process is, 
\begin{eqnarray}
g_{h_{1}tt} &=& -\dfrac{M_{t}}{v} \cos\alpha \,, \nn \\
g_{h_{2}tt} &=& \dfrac{M_{t}}{v} \sin\alpha \,, \nn \\
M_{tt} &=& \dfrac{g_{h_{1}tt}\,
\,g_{h_{1}\phi_{DM}^{\dagger}\phi_{DM}}}{(s-M_{h_1}^{2}) +
i M_{h_1} \Gamma_{h_1}} + \dfrac{g_{h_{2}tt}
\,\,g_{h_{2}\phi_{DM}^{\dagger}\phi_{DM}}}{(s-M_{h_2}^{2}) +
i M_{h_2} \Gamma_{h_2}}, \nn \\
\sigma_{{\rm t}{\rm \bar{t}} \rightarrow \phi_{DM}^{\dagger}\phi_{DM}} &=&
\dfrac{1}{32 \pi s n_{c}}\,\,(s - 4M_{t}^{2})\,\,
\sqrt{\dfrac{s - 4M_{DM}^{2}}{s - 4M_{t}^{2}}}
\,\,\,|M_{tt}|^{2}\,,
\end{eqnarray}
where $M_{t}$ is the mass of the top quark and $n_{c}=3$ is its
color charge.

\item \underline{$N_{1}$\,$N_{j}$\,$\rightarrow
\phi_{DM}^{\dagger}\,\phi_{DM}\,\,(j=2,\,3)$}\,:
\newline
\newline
The annihilation of $N_1$ and $N_j$ to $\phi_{DM}^{\dagger}\,\phi_{DM}$
has two $s$\,channel diagrams mediated by $h_1$ and $h_2$ respectively.
The corresponding expression of the cross section is, 
\begin{eqnarray}
g_{h_{1}N_{1}N_{2\,(3)}} &=& \sqrt{2}\,h_{e\mu\,(\tau)}\sin\alpha \,, \nn \\
g_{h_{2}N_{1}N_{2\,(3)}} &=& \sqrt{2}\,h_{e\mu\,(\tau)}\cos\alpha \,, \nn \\
M_{N_{1}N_{j}} &=& \dfrac{g_{h_{1}N_{1}N_{j}}\,
\,g_{h_{1}\phi_{DM}^{\dagger}\phi_{DM}}}{(s-M_{h_1}^{2}) +
i M_{h_1} \Gamma_{h_1}} + \dfrac{g_{h_{2}N_{1}N_{j}}
\,\,g_{h_{2}\phi_{DM}^{\dagger}\phi_{DM}}}{(s-M_{h_2}^{2}) +
i M_{h_2} \Gamma_{h_2}}\,, \nn \\
\sigma_{N_{1}N_{j} \rightarrow \phi_{DM}^{\dagger}\phi_{DM}} &=&
\dfrac{(s - (M_{N_1} + M_{N_j})^{2})}{32 \pi s}\times\nn \\
&&\sqrt{\dfrac{s(s - 4M_{DM}^{2})}
{(s-(M_{N_j}+M_{N_1})^{2})(s - (M_{N_j} - M_{N_1})^{2})}}
\,\,\,|M_{N_{1}N_{j}}|^{2}\,. \nn \\
\end{eqnarray}
\item \underline{$N_{j}$\,$N_{j}$\,
$\rightarrow \phi_{DM}^{\dagger}\,\phi_{DM}\,\,(j=2,\,3)$}\,:
\newline
\newline
Unlike the previous cases, here the annihilation of
$N_{j}$\,$N_{j}$ occurs only through a $s$\,channel
process mediated by the extra neutral gauge boson $\zmt$. The
expression of the cross section is,
\begin{eqnarray}
&&\sigma_{N_{j}N_{j} \rightarrow \phi_{DM}^{\dagger}\phi_{DM}} =
\dfrac{g_{\mu\tau}^{4}n_{\mu\tau}^{2}}{192 \pi s} \,
\sqrt{\dfrac{s-4M_{DM}^{2}}{s-4M_{N_j}^{2}}}
\dfrac{(s-4M_{DM}^{2})(s-4M_{N_j}^{2})}{[(s - M_{Z_{\mu\tau}}^{2})^{2} +
\Gamma_{Z_{\mu\tau}}^{2}M_{Z_{\mu\tau}}^{2}]}\,.
\label{n1n2goingphidm}
\end{eqnarray}
\end{itemize} 

\section{Expressions of decay widths of $h_2$, $h_1$ and $\zmt$}
\label{App:AppendixB}

\underline{\bf Total decay width of $h_2$ :}\\
Decay width for a particular process is generally
calculated in the rest frame of the corresponding decaying particle.
In the present work, the BSM Higgs $h_2$ can decay
to all the SM particles and also to other BSM particles like $\zmt$,
$\phi_{DM}$ and $N_j$. Here, we have given the expressions
of all partial decay widths as well as the total decay
width of $h_2$.
\begin{itemize}
\item \underline{$h_2$ $\rightarrow VV$} ($V= W^\pm, Z$):
\newline
The width of this decay process is given as follows
\begin{eqnarray}
g_{h_{2}VV} &=& -\dfrac{2M_{V}^{2}}{v}\,\sin\alpha\,,\nn \\
\Gamma(h_2 \rightarrow VV) &=&
\dfrac{M_{h_2}^{3}\,g_{h_{2}VV}^{2}}
{64\,\pi M_{V}^{4}\,S_{V}}
\sqrt{1 - \dfrac{4 M_{V}^{2}}{M_{h_2}^{2}}}\,\,
\left(1 - \dfrac{4 M_{V}^{2}}{M_{h_2}^{2}}
+ \dfrac{12M_{V}^{4}}{M_{h_2}^{4}}\right)\,.
\end{eqnarray}
In the above decay expression $S_{V}$ is the statistical
factor. It is 1 for $W^\pm$ boson and 2 for $Z$ boson.
\\
\item \underline{$h_2$ $\rightarrow \zmt\zmt$}:
\newline
The expression of decay width of $h_2$ $\rightarrow \zmt\zmt$ is,
\begin{eqnarray}
g_{h_{2}Z_{\mu\tau}Z_{\mu\tau}} &=&\dfrac{2\,M_{Z_{\mu\tau}}^{2}}
{v_{\mu\tau}}\,\cos\alpha\,,\nn \\
\Gamma(h_2 \rightarrow Z_{\mu\tau}Z_{\mu\tau}) &=&
\dfrac{M_{h_2}^{3}\,g_{h_{2}Z_{\mu\tau}Z_{\mu\tau}}^{2}}{128\,\pi
M_{Z_{\mu\tau}}^{4}}
\sqrt{1 - \dfrac{4 M_{Z_{\mu\tau}}^{2}}{M_{h_2}^{2}}}\,\,
\left(1 - \dfrac{4 M_{Z_{\mu\tau}}^{2}}{M_{h_2}^{2}}
+ \dfrac{12M_{Z_{\mu\tau}}^{4}}{M_{h_2}^{4}}\right)\,.
\end{eqnarray}

\item \underline{$h_{2}$ $\rightarrow$ $h_1$\,$h_1$}:
\newline
The expression of the decay width is,
\begin{eqnarray}
&&\Gamma({h_{2}} \rightarrow {h_1}{h_1}) = 
\dfrac{g_{h_{1}h_{1}h_{2}}^{2}}{16\,\pi\,M_{h_2}S}
\,\sqrt{1 - \dfrac{4M_{h_1}^{2}}{M_{h_2}^{2}}}\,.
\end{eqnarray}
Here the statistical factor $S=2$. The expression of the
coupling constant $g_{h_{1}h_{1}h_{2}}$ is given in
Eq.~(\ref{h2h1h1}). 
\item \underline{$h_{2}$ $\rightarrow$ $\phi_{DM}^{\dagger}$
\,$\phi_{DM}$}:
\newline
Similarly, the expression of the decay width of
$h_{2}$ $\rightarrow$ $\phi_{DM}^{\dagger}\,\phi_{DM}$ can be written
as,
\begin{eqnarray}
\Gamma({h_{2}} \rightarrow \phi_{DM}^{\dagger}\phi_{DM}) &=& 
\dfrac{g_{h_{2}\phi_{DM}^{\dagger}\phi_{DM}}^{2}}{16\,\pi\,M_{h_2}}
\,\sqrt{1 - \dfrac{4 M_{DM}^{2}}{M_{h_2}^{2}}}\,,
\end{eqnarray}
where the expression of the coupling $g_{h_{2}\phi_{DM}^{\dagger}
\phi_{DM}}$ is given in Eq.~(\ref{dm_vertex}).
\item \underline{$h_{2}$ $\rightarrow$ $f\bar{f}$}:
\newline
Here $f$ represents all SM fermions and the corresponding
expression decay width is,
\begin{eqnarray}
g_{h_{2}ff} &=& \dfrac{M_{f}}{v}\,\sin\alpha \,,\nn \\
\Gamma(h_{2} \rightarrow {\rm f} \,\bar{\rm f})
&=& \dfrac{n_{c}\, M_{h_{2}}\, g_{h_{2}ff}}{8\pi}
\,\left(1 - \dfrac{4 M_{f}^{2}}{M_{h_2}^{2}}\right)
^{\frac{3}{2}}\,,
\end{eqnarray}
where $n_c$ is the color charge, it is 1 for leptons and 3 for quarks.

\item \underline{$h_{2}$ $\rightarrow$ $N_1$\,${N_j}\,\,
(j=2,\,3)$}:
\newline
The partial decay width of $h_2$ to $N_1$\,${N_j}$ is given as,
\begin{eqnarray}
g_{h_{2} N_1 N_{2\,(3)}} &=& \sqrt{2}\,\sin\alpha \,h_{e\,\mu(\tau)} \,,\nn \\
\Gamma(h_{2} \rightarrow {N_1} \,{N_j}) &=& \dfrac{M_{h_{2}}\,
g_{h_{2} N_1 N_j}}{8\pi}
\,\left(1 - \dfrac{(M_{N_1} + M_{N_j})^{2}}
{M_{h_2}^{2}}\right)^{\frac{3}{2}}\,.
\end{eqnarray}
\end{itemize}
Finally, the total decay width of the BSM Higgs $h_2$
is thus the sum of all partial decay
widths \footnote{Here we have taken only those decay modes
of $h_2$ which occur in tree level.} mentioned above , which is
\begin{eqnarray}
&&\Gamma_{h_2} = \sum_{V=W,Z}\Gamma(h_2 \rightarrow VV) +
\Gamma(h_2 \rightarrow Z_{\mu\tau}Z_{\mu\tau}) +
\Gamma({h_{2}} \rightarrow {h_1}{h_1})
+ \Gamma({h_{2}} \rightarrow \phi_{DM}^{\dagger}\phi_{DM}) \nn \\
&&~~~~~~~~~
 + \sum_{f}\Gamma(h_{2} \rightarrow {f} \,\bar{f})
+\sum_{j = 2,\,3}\Gamma(h_{2} \rightarrow {N_1} \,{N_j})\,.
\end{eqnarray}
\underline{\bf Decay width of h$_1$}:\\
Besides decaying to SM particles, $h_1$
has extra decay modes to a pair of $\zmt$ and $\phi_{DM}$ respectively.
The expressions of these decay channels are,
\begin{eqnarray}
g_{h_{1}Z_{\mu\tau}Z_{\mu\tau}} &=& \dfrac{2\,M_{Z_{\mu\tau}}^{2}}
{v_{\mu\tau}}\,\sin\alpha\,,\nn \\
\Gamma(h_1 \rightarrow Z_{\mu\tau}Z_{\mu\tau}) &=&
\dfrac{M_{h_1}^{3}\,g_{h_{1}Z_{\mu\tau}Z_{\mu\tau}}^{2}}{128\,\pi
M_{Z_{\mu\tau}}^{4}}
\sqrt{1 - \dfrac{4 M_{Z_{\mu\tau}}^{2}}{M_{h_1}^{2}}}\,\,
\left(1 - \dfrac{4 M_{Z_{\mu\tau}}^{2}}{M_{h_1}^{2}}
+ \dfrac{12M_{Z_{\mu\tau}}^{4}}{M_{h_1}^{4}}\right)\,,\\
\Gamma({h_{1}} \rightarrow \phi_{DM}^{\dagger}\phi_{DM}) &=& 
\dfrac{g_{h_{1}\phi_{DM}^{\dagger}\phi_{DM}}^{2}}{32\,\pi\,M_{h_1}}
\,\sqrt{1 - \dfrac{4 M_{DM}^{2}}{M_{h_1}^{2}}}\,,
\end{eqnarray}
where the expression of $g_{h_{1}\phi_{DM}^{\dagger}\phi_{DM}}$
is given in Eq.~(\ref{dm_vertex}). Therefore, total decay width
of the SM-like Higgs $h_1$ can be written as,
\begin{eqnarray}
\Gamma_{h_1} = \cos^{2}\alpha \,\Gamma_{SM} +
\Gamma(h_1 \rightarrow Z_{\mu\tau}Z_{\mu\tau}) +
\Gamma({h_{1}} \rightarrow \phi_{DM}^{\dagger}\phi_{DM})\,\,.
\end{eqnarray}
\underline{\bf Decay width of Z$_{\mu\tau}$}:\\
Since in this work we have considered the low mass of $\zmt$
($\mzmt\sim 100$ MeV) and also it has no couplings to quarks, hence
it can only decay to neutrinos.
Therefore expression of total decay width of $\zmt$ is given by
\begin{eqnarray}
\Gamma_{Z_{\mu\tau}} = \sum_{j = 1,2,3}
\dfrac{g_{\mu\tau}^{2} M_{Z_{\mu\tau}}}
{96\,\pi}
\left(1 - \dfrac{4 M_{\nu_j}^{2}}{M_{Z_{\mu\tau}}}\right)^{\dfrac{3}{2}}\,.
\end{eqnarray}  

\end{document}